\documentclass[10pt,aps,twocolumn,secnumarabic,amssymb, nobibnotes,  prx,superscriptaddress,floatfix]{revtex4-2}

\usepackage[normalem]{ulem}
\usepackage{todonotes}
\usepackage{amsmath}
\usepackage{graphicx}
\usepackage[colorlinks=true, allcolors=blue]{hyperref}
\usepackage{tabularx}
\usepackage{ragged2e} 

\usepackage{physics}
\usepackage{tikz}
\usepackage{pgfplots}
\pgfplotsset{compat=1.18}
\usetikzlibrary{calc, positioning, pgfplots.fillbetween}

\begin{document}

\title{Enabling full localization of qubits and gates with a multi-mode coupler}

\author{Zhongyi Jiang}
\affiliation{Institute for Quantum Information, RWTH Aachen University, D-52056 Aachen, Germany}
\affiliation{Peter Gr\"unberg Institute PGI-12, Forschungszentrum J\"ulich, J\"ulich 52425, Germany}

\author{Simon Geisert}
\affiliation{IQMT, Karlsruhe Institute of Technology, 76344 Eggenstein-Leopoldshafen, Germany}

\author{Sören Ihssen}
\affiliation{IQMT, Karlsruhe Institute of Technology, 76344 Eggenstein-Leopoldshafen, Germany}

\author{Ioan M. Pop}
\affiliation{IQMT, Karlsruhe Institute of Technology, 76344 Eggenstein-Leopoldshafen, Germany}
\affiliation{PHI, Karlsruhe Institute of Technology, 76131 Karlsruhe, Germany}
\affiliation{Physics Institute 1, Stuttgart University, 70569 Stuttgart, Germany}

 \author{Mohammad H. Ansari}
\affiliation{Institute for Quantum Information, RWTH Aachen University, D-52056 Aachen, Germany}
\affiliation{Peter Gr\"unberg Institute PGI-12, Forschungszentrum J\"ulich, J\"ulich 52425, Germany}

\begin{abstract}

Tunable couplers are a key building block of superconducting quantum processors, enabling high on–off ratios for two-qubit entangling interactions. While qubit–qubit interaction can be turned off, residual wavefunctions delocalize single-qubit excitations over the device, yielding weak effective couplings that manifest as unintended crosstalk. Moreover, conventional single-mode couplers lack independent control over interactions in the one- and two-excitation manifolds, leading to unitary errors such as leakage during gate operations. Here, we propose a multi-mode tunable coupler that enforces complete localization, yielding near-perfect qubit isolation at the decoupled point. We further show that the additional degrees of freedom in the coupler enable independent and nonlinear control of effective interactions across distinct excitation manifolds, with large on/off ratios. This architecture provides a new route toward the next generation of couplers for scalable and high-fidelity gate operations in superconducting quantum processors.

\end{abstract}

\date{\today}

\maketitle

\section{Introduction}
Superconducting quantum information processing systems have emerged as one of the most promising platforms for realizing large-scale quantum computing. Several experiments have already demonstrated that state-of-the-art superconducting processors can outperform classical supercomputers in certain tasks~\cite{Arute_2019,Kim_2023}. The preparation of entangled states on these processors is particularly vulnerable to errors. Effective denoising requires meticulous characterization of noise sources and the development of protocols that mitigate these disturbances without modifying the quantum state.  Advancements in state denoising mark one of the critical milestones toward realizing scalable and fault-tolerant quantum computing. There are deterministic protocols for addressing denoising such as quantum error correction \cite{Bravyi_Cross_Gambetta_Maslov_Rall_Yoder_2024} and error mitigation techniques, such as quantum autoencoders \cite{Pazem_2025}. Therefore, achieving fault-tolerant quantum computing has become the next central objective of the community~\cite{PhysRevA.52.R2493,ai2024quantum}. To implement quantum error correction, the physical qubit and gate errors must be reduced below a certain threshold~\cite{PhysRevA.52.R2493,Fowler_2012,Bravyi_Cross_Gambetta_Maslov_Rall_Yoder_2024,mohseni2025buildquantumsupercomputerscaling}.

Although single-qubit gate errors can already be suppressed well below $0.1\%$, sometimes reaching $0.01\%$ or even lower~\cite{PRXQuantum.5.040342}, two-qubit gate errors cannot yet be consistently reduced below $0.1\%$~\cite{PhysRevX.14.041050,Ding_2023}. Consequently, from a hardware perspective, the design and control of high-fidelity two-qubit gates remains one of the key challenges~\cite{ai2024quantum,mohseni2025buildquantumsupercomputerscaling}. 

In superconducting systems, the most common approach to controlling two-qubit interactions is through a tunable coupler, typically realized with a single tunable mode that mediates the interaction~\cite{PhysRevLett.113.220502}. Such couplers enable versatile gate schemes~\cite{McKay_2016,Yan_2018,Liang_2023}. Recent lattice Hamiltonian techniques \cite{Xu_2024} and surface-code QPU Hamiltonian \cite{xu2025surfacecodehardwarehamiltonian} show the importance of tunable couplers to suppress unwanted two-body and three-body interactions \cite{xu2025paritycrossresonancemultiqubitgate} in the soft qubit decoupling regime, which allows the achievement of large on-off ratios~\cite{PhysRevLett.113.220502}. 

Despite these advantages, the single-mode tunable coupler (SMC) design suffers from intrinsic limitations. A primary issue is that qubits remain partially delocalized, even when the coupler is tuned to the nominal “interaction-off” point~\cite{Berke_2022,Heunisch_2023}. Qubit wavefunctions inevitably leak into nearby qubits, generating crosstalk and degrading gate fidelity~\cite{Lange_2026}. Another limitation is the lack of independent control over qubit–qubit interactions in the one-excitation and two-excitation subspaces. Because the interaction is mediated exclusively by a single tunable mode, both subspaces are activated simultaneously during a two-qubit gate~\cite{xu2025surfacecodehardwarehamiltonian,PhysRevApplied.19.024057}, leading to leakage and unitary errors. 

\begin{figure}[htbp]
  \centering

    \includegraphics[width=0.96\linewidth]{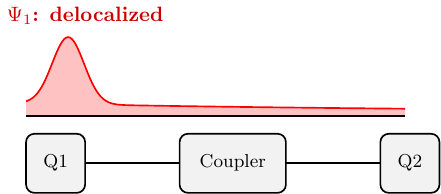}
    \put(-230,110){\textbf{(a)}} 
\hfil
\vspace{0.1cm}
\hfill
\centering
    \includegraphics[width=0.96\linewidth]{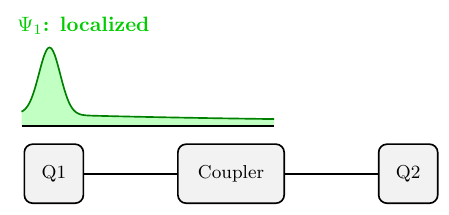}
    \put(-230,115){\textbf{(b)}} 

\hfill
    \includegraphics[width=0.48\textwidth]{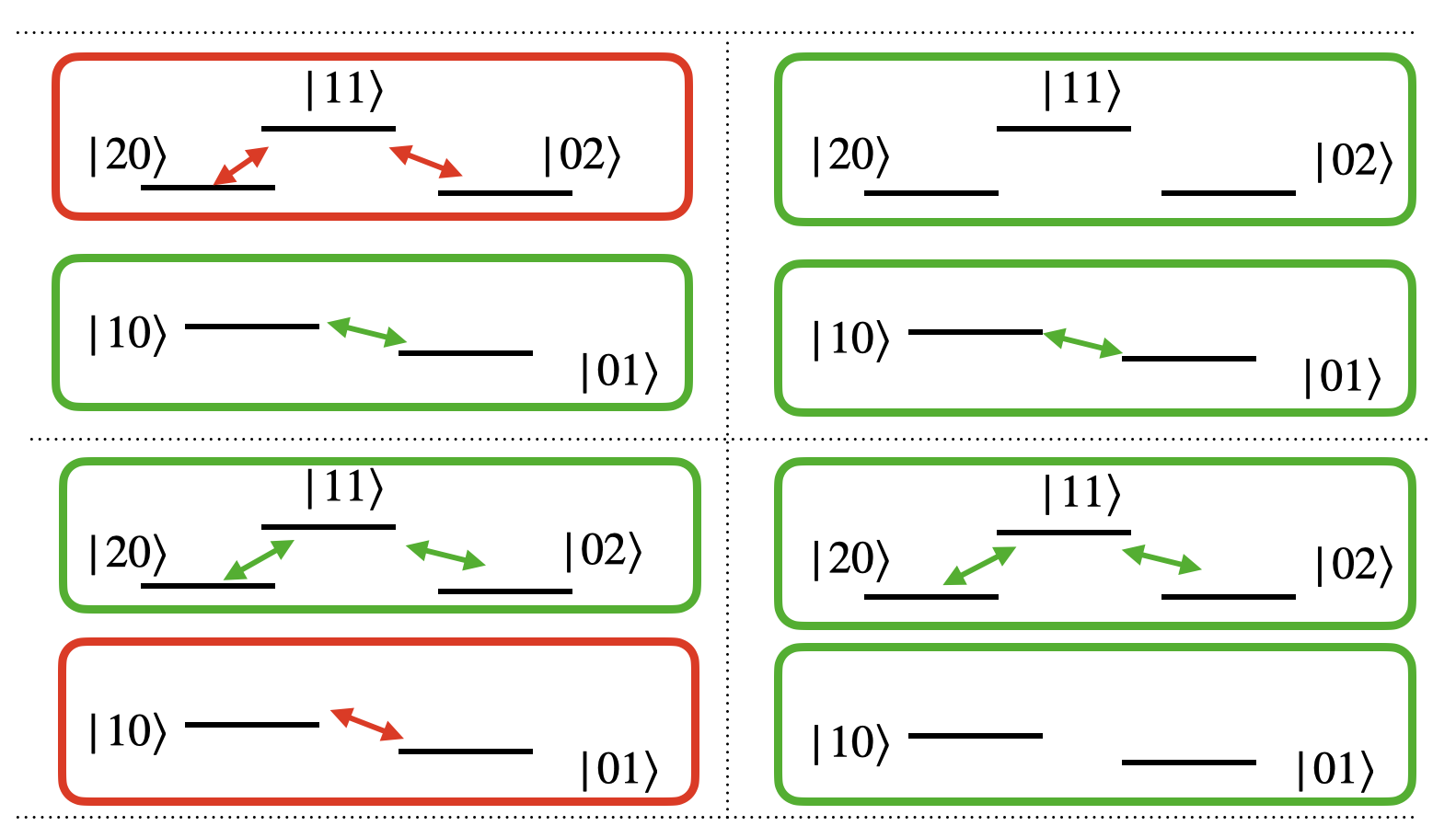}\put(-230,135){\textbf {(c)}}

  \caption{\textbf{Comparison between a single-mode coupler~(SMC) and a two-mode coupler~(TMC).} In (a) and (b), we compare wavefunction delocalization when interactions are turned off. We draw the spatial locations of qubit modes and coupler modes. For an SMC in (a), there is always-on delocalization of wavefunctions from one qubit to the other, whereas for a TMC in (b), this delocalization can be confined within the coupler. In (c), we compare the J coupling strength in the one-excitation subspace and the two-excitation subspace. The coupler states are dropped for simplicity. In the SMC case~(left), residual interaction in the two-excitation subspace can not be independently controlled. In the TMC case~(right), interactions in the two-excitation subspace can be separately turned on. In the case of $i$SWAP gates~(top), interactions in the two-excitation manifold remain zero. In the case of CPHASE gates~(bottom), interactions in the one-excitation manifold remain zero. }
  \label{fig:motivation}
\end{figure}

For example, consider implementing an iSWAP gate between two qubits. This requires turning on the interaction between $|01\rangle$ and $|10\rangle$, denoted as $J_{00}$. However, due to the absence of independent control, interactions between $|11\rangle$ and $|02\rangle$, as well as between $|11\rangle$ and $|20\rangle$, are turned on with comparable strength. Although the dynamics of $|11\rangle$ is suppressed relative to the $i$SWAP transition due to larger detunings, these residual interactions still induce small oscillations and dispersive shifts of $|11\rangle$, which manifest themselves as leakage and undesired phase errors $ZZ$. The residual interactions can also affect CPHASE gates by inducing unwanted transitions between $|01\rangle$ and $|10\rangle$. These scenarios are illustrated in FIG.~\ref{fig:motivation}. 

A similar issue arises in idle mode: even when no interaction is desired, residual couplings often persist. This occurs even if the effective qubit–qubit interaction strength within the computational subspace is tuned to zero. In practice, computationally noninteracting qubits may still couple through higher-energy, noncomputational states, giving rise to the so-called \emph{stray couplings}~\cite{PhysRevLett.124.230503,PhysRevApplied.15.064074}.

To address the limitations of single-mode couplers (SMCs), a recent design introduced a multi-mode coupler that allows a qubit to couple to a left-handed metamaterial ring resonator \cite{PRXQuantum.5.020325}. Here, we propose a two-mode tunable coupler (TMC) and generalize it to multiqubit-multimode architectures. Couplers with multiple modes have already been proposed and experimentally demonstrated in various contexts \cite{Goto_2022,PhysRevX.14.041050,Mundada_2019,Moskalenko_2021,PRXQuantum.5.020325,khoshnegar2014toward}. However, they do not focus on reducing delocalization or independent coupling control. In our TMC design, we address both delocalization and independent control of interactions. The core of our TMC is two tunable modes and tunable mode-mode interaction. With fully tunable mode frequencies and coupling, qubits can be decoupled with full localization. The wavefunctions of qubits are limited to individual qubits and associated coupler modes, which we name localized decoupling. 

We emphasize that our goal is not to completely localize the qubit wavefunctions within the qubit modes. Such complete localization is neither necessary nor possible unless the qubits are fully isolated. Instead, we allow for delocalization between the qubits and the coupler modes, while only requiring that the qubit wavefunctions remain localized with respect to each other. The comparison of delocalization and localization is illustrated in FIG.~\ref{fig:motivation} (a) and (b). Interactions in different excitation subspaces can be switched on independently. This can be used to implement pure $i$SWAP operation or pure CPHASE operation without interference from one another. It can also be used to implement $i$SWAP and CPHASE simultaneously, as in continuous fSim gates\cite{Foxen_2020,jiang2024concurrentfermionicsimulationgate}. 
The rest of the paper is organized as follows.

We first lay out the general framework of our theory of multi-mode couplers and a new method for evaluating effective couplings in Sec.~\ref{sec:general theo}. We then study a concrete case of TMC and show how fully localized decoupling can be achieved in Sect.~\ref{sec:full_localization}. Circuit designs for TMC are also proposed for experiments. In Sect.~\ref{sec: selective J}, we show analytical and numerical evidence that effective couplings in different excitation manifolds can be selectively controlled. Finally, in Sect.~\ref{sec:modular QPU}, we present a vision for modular chip design based on multi-mode couplers, aimed at enabling the scalable development of superconducting quantum processors.

\section{ Theory  of Interacting Couplers}\label{sec:general theo}

The circuits considered in this work incorporate more than one coupling mode, and these modes can interact with each other. In particular, the coupling strength between different coupler modes is a tunable physical parameter that can be switched on or off experimentally.  

We consider a system of $N$ qubits coupled to $M$ coupler modes. Depending on the design, the couplers may mediate all-to-all qubit interactions or be localized as lumped elements that provide pairwise qubit coupling. The general Hamiltonian of the system can be written as
\begin{eqnarray}
    && H = H_C + H_{QC} + H_Q,  \nonumber \\
    && \mathrm{and }\ \  H_C =\sum_{i}^M H_{c_i} + \sum_{i \neq j}^M H_{c_i c_j}(\lambda),
    \label{eq.H}
\end{eqnarray}
where $H_Q=\sum_{i=1}^N H_{q_i}$ describes the qubits, $H_C$ the coupler modes, and $H_{QC}=\sum_i^N \sum_{j=1}^M H_{q_ic_j}$ their mutual interaction.  The coupler Hamiltonian $H_C$ is naturally separated into two parts: (i) the diagonal terms $H_{c_i}$, corresponding to all individual modes $c_i$, and (ii) the off-diagonal mode–mode interaction terms $H_{c_i c_j}(\lambda)$, representing the coupling between modes $i$ and $j$, with $i,j \in \{1,2,\dots,M\}$ and $\lambda$ denoting a physical knob for tuning the coupling strength between the two modes.   In analogy to the standard notation for qubits \cite{PhysRevB.100.024509,RevModPhys.93.025005}, diagonal coupling terms describe the couplers on a bare-mode basis, while off-diagonal terms account for the renormalization induced by mode–mode interactions.

\begin{figure}[h]
\includegraphics[width=0.3\textwidth]{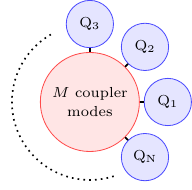}\put(-180,110){\textbf {(a)}}\\ \includegraphics[width=0.46\textwidth] {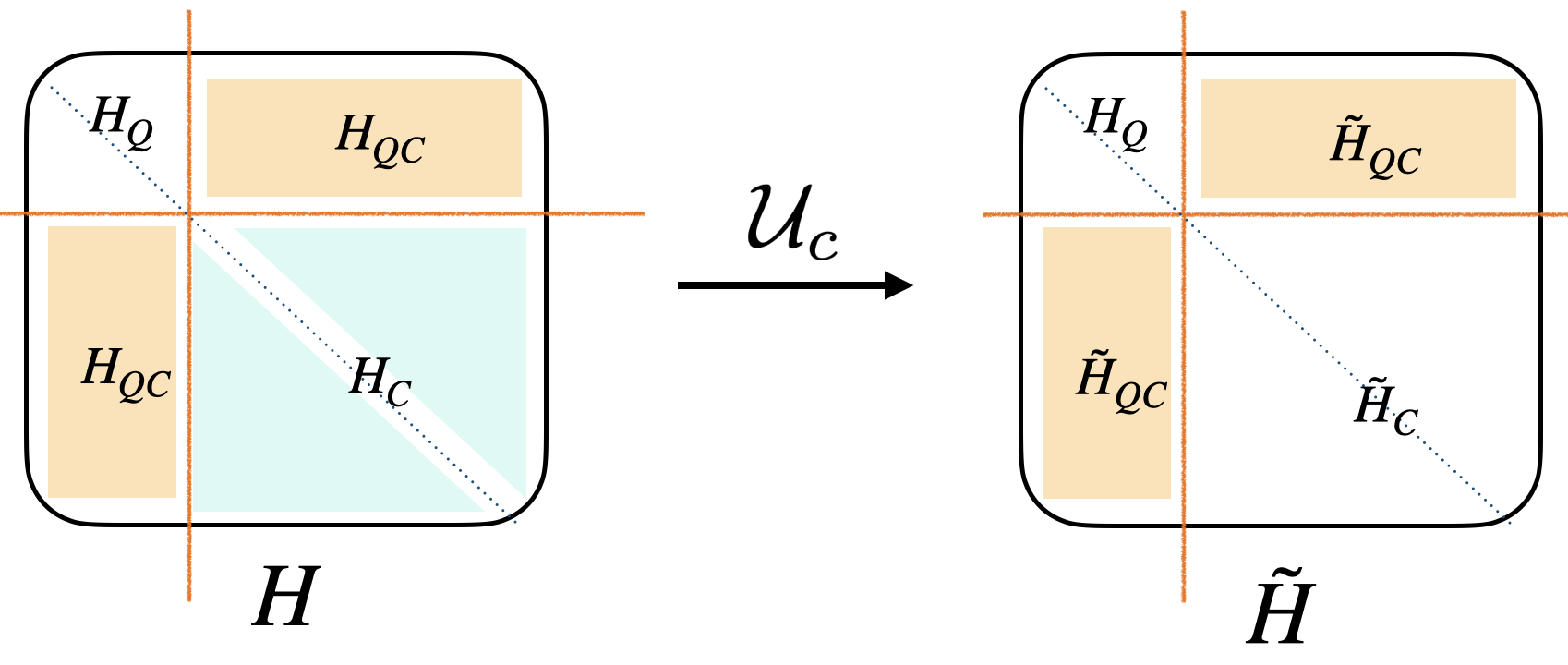}
\put(-220,100){\textbf {(b)}}
\caption{\textbf{Implementing qubit-coupler wavefunction localization.} (a) Schematic of a system with $N$ qubits coupled through $M$ coupler modes. (b) Corresponding Hamiltonian decomposition (left), as in Eq.~(\ref{eq.H}), into qubit and coupler subspaces and diagonalizing the coupler using the transformation $\mathcal{U}_c$. The qubit subspace will stay invariant however qubits are affected via the variation of qubit-coupler interactions.}
        \label{fig:H full}
\end{figure}

The qubit Hamiltonian $H_Q$ is, in general, non-diagonal: direct qubit–qubit couplings appear as small off-diagonal terms. Although typically weak, these couplings cannot be neglected, as they shift the effective qubit–qubit interaction points~\cite{Xu_2024,xu2025surfacecodehardwarehamiltonian}. This shift can be compensated or made bigger in the presence of (tunable) couplers.  
For clarity, our analytical derivations and schematic diagrams omit explicit direct qubit–qubit couplings. However, our numerical simulations fully include them, demonstrating that the proposed multi-mode coupler architecture can successfully compensate for such residual interactions.

A key feature of our model is the mode–mode coupling, denoted $H_{cc}(\lambda)$. This term plays a central role because we consider couplers whose modes are strongly hybridized and whose mode–mode interactions are tunable. In previously proposed designs, the situation is markedly different: either (i)~the modes are weakly coupled, so $H_{cc}\approx 0$ is negligible~\cite{Goto_2022,PhysRevX.14.041050,Mundada_2019,Moskalenko_2021,PRXQuantum.5.020325}, or (ii)~the mode–mode coupling $H_{cc}(\lambda)$ is a fixed value over a large domain of $\lambda$, thus can be removed by re-framing the Hamiltonian into a diagonal form of the coupler Hamiltonian \cite{PRXQuantum.5.020325}.   In contrast, our design combines strong mode hybridization with \emph{tunable} mode–mode coupling, making $H_{cc}$ essential. As we will show, this term is also a crucial ingredient for realizing localized decoupling within the qubit network.  A schematic qubit setup and the Hamiltonians are shown in FIG. \ref{fig:H full}.

Next, we demonstrate how the coupling between qubits can be switched on or off by selectively controlling the mode–mode interactions within the coupler.

\subsection{Locally Confined Qubit–Coupler Blocks}

The central idea of our qubit–coupler confinement scheme is to achieve \emph{complete decoupling} by partitioning the circuit of $N$ qubits and $M$ couplers/coupling modes into $N$ independent subsystems, each containing a single qubit and one or more coupler modes. This confinement is supposed to suppress all interactions between blocks, not only within the computational subspace but across the entire Hilbert space. Moreover, the decoupling achieved is localized. Qubit wavefunctions are confined within its own block without any leakage to other qubits.

We show that such confinement can be realized by tuning the parameter $\lambda$ in the mode–mode coupling $H_{cc}(\lambda)$. To demonstrate this, we first diagonalize the coupler Hamiltonian using a unitary transformation $\mathcal{U}_c$ acting solely on the coupler degrees of freedom. Since the mode–mode coupling is tunable, $\mathcal{U}_c$ itself can be externally controlled via $\lambda$, for instance, through tunable flux qubits or superconducting quantum interference devices.

After the transformation, the total Hamiltonian takes the following form:
\begin{equation}
    \tilde{H}=\mathcal{U}^\dagger _c H \mathcal{U}_c 
\label{eq. UcHUc}
\end{equation}
The operator $\mathcal{U}_c$ depends on the physical tuning parameters $\lambda$ and must act on the Hamiltonian in the \emph{full} Hilbert space, prior to any truncation to the computational subspace. We note that the external control $\lambda$ in practice can consist of multiple parameters, $\lambda=(\lambda_1,...,\lambda_n)$. For simplicity, we use the symbol $\lambda$ to represent all external controls. Since the transformation involves no qubit degrees of freedom, Eq.~(\ref{eq. UcHUc}) yields the transformed total Hamiltonian  
\begin{equation}
    \tilde{H} = H_Q + \mathcal{U}^\dagger_c(\lambda) H_C \mathcal{U}_c(\lambda) + \mathcal{U}^\dagger_c(\lambda) H_{QC}(\lambda) \mathcal{U}_c(\lambda),
\end{equation}
which can be written compactly as  
\begin{equation}
    \tilde{H} = H_Q + \tilde{H}_C(\lambda) + \tilde{H}_{QC}(\lambda).
\end{equation}

\begin{figure}[ht!]
    \centering
    \includegraphics[width=0.46\textwidth]{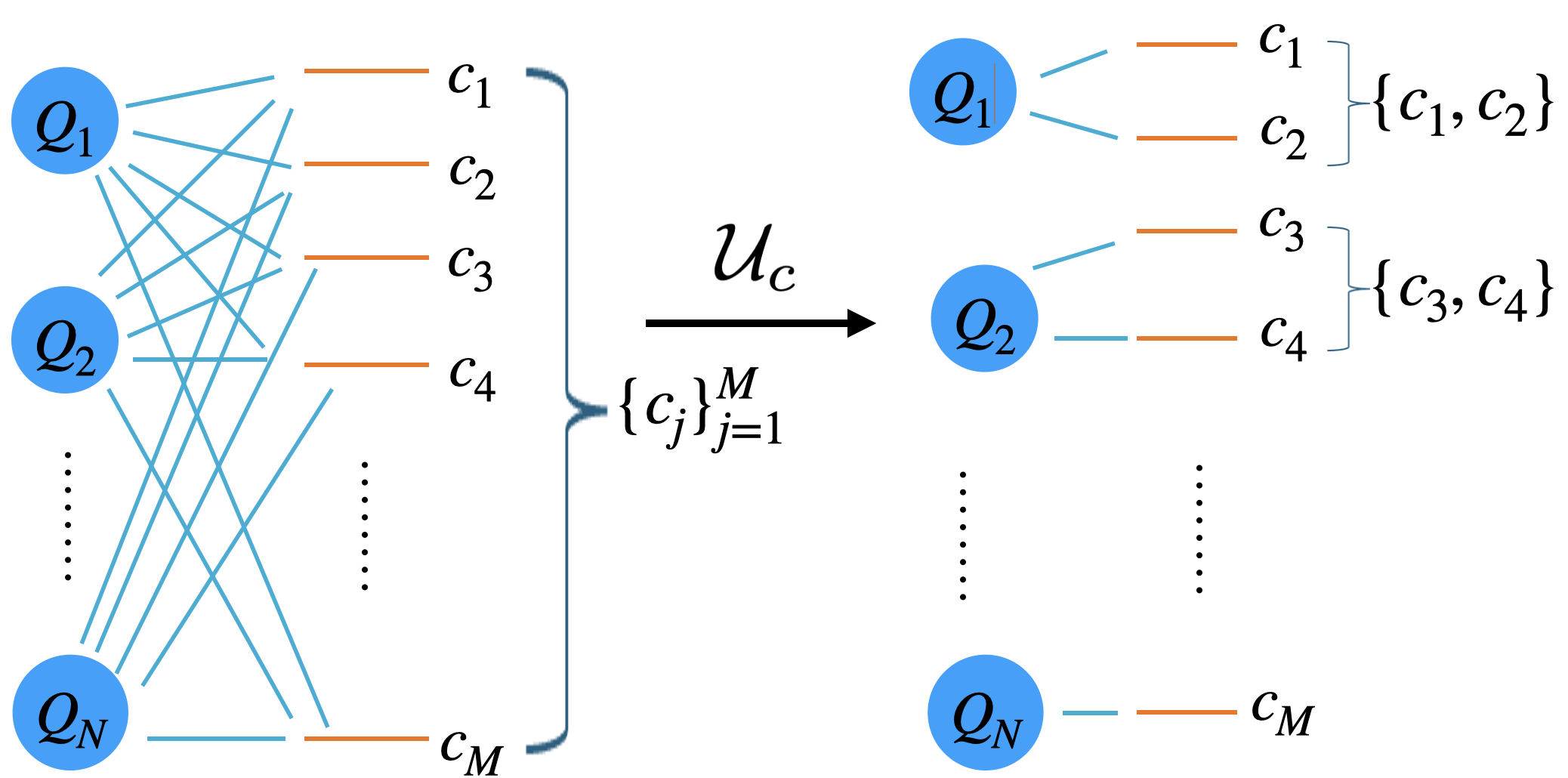}
        \caption{\textbf{Partitioning of coupler modes.}  In the example shown, qubit $1$ couples to ${c_1,c_2}$ ($m_1=2$) and qubit $2$ couples to ${c_3,c_4}$ ($m_2=2$), yielding $N$ non-interacting qubit-coupler blocks. The ordering of the dressed modes after $\mathcal{U}_c$ need not match the original indexing.}
        \label{fig:partition}
\end{figure}

FIG.~\ref{fig:H full}(b) illustrates this transformation in a matrix representation.
After applying the coupler transformation $\mathcal{U}c(\lambda)$, the $M$ modes $\{c_1,c_2,\ldots,c_M\}\equiv\{c_j\}_{j=1}^M$ are grouped into $N$ disjoint subgroups $\mathcal{C}_i\equiv \{c_{i_k}\}_{k=1}^{m_i}$  (one set per qubit, with $\sum_{i=1}^N m_i=M$), so that in $\tilde{H}_{QC}$ each term couples one qubit only to the coupler modes of its associated subgroup. An illustrative example is shown in FIG.~\ref{fig:partition}.


The interaction Hamiltonian $\tilde{H}_{QC}$ can be split into the following form:
\begin{equation}
\tilde{H}_{QC}(\lambda)=\sum_{i=1}^N H_{Q_i\mathcal{C}i}(\lambda)
\label{eq.HQC}
\end{equation}
where $H_{Q_i\mathcal{C}_i}(\lambda)=\sum_{c\in \mathcal{C}_i} H_{q_i c}(\lambda)$ is constructed from operators of qubit $i$ and the coupler modes of the corresponding group $\mathcal{C}_i$.

The diagonalized coupling parts of the Hamiltonian in Eq. (\ref{eq. UcHUc}), i.e. $\tilde{H}_{C}$, can be grouped accordingly into the modes associated with the subgroups $\mathcal{C}_i$ so that 
\begin{equation}
\tilde{H}_{C}(\lambda)=\sum_{i=1}^N H_{\mathcal{C}_i}(\lambda)
\label{eq.Hc}    
\end{equation}
with $H_{\mathcal{C}_i}(\lambda) \equiv \sum_{j_i\in \mathcal{C}_i} H_{c_{j_i}}(\lambda)$. Thus, the full Hamiltonian can be rewritten as a sum of $N$ commuting Hamiltonians supported on disjoint qubit–coupler groups. 

\begin{figure}[ht!]
     \centering
     \includegraphics[width=0.46\textwidth]{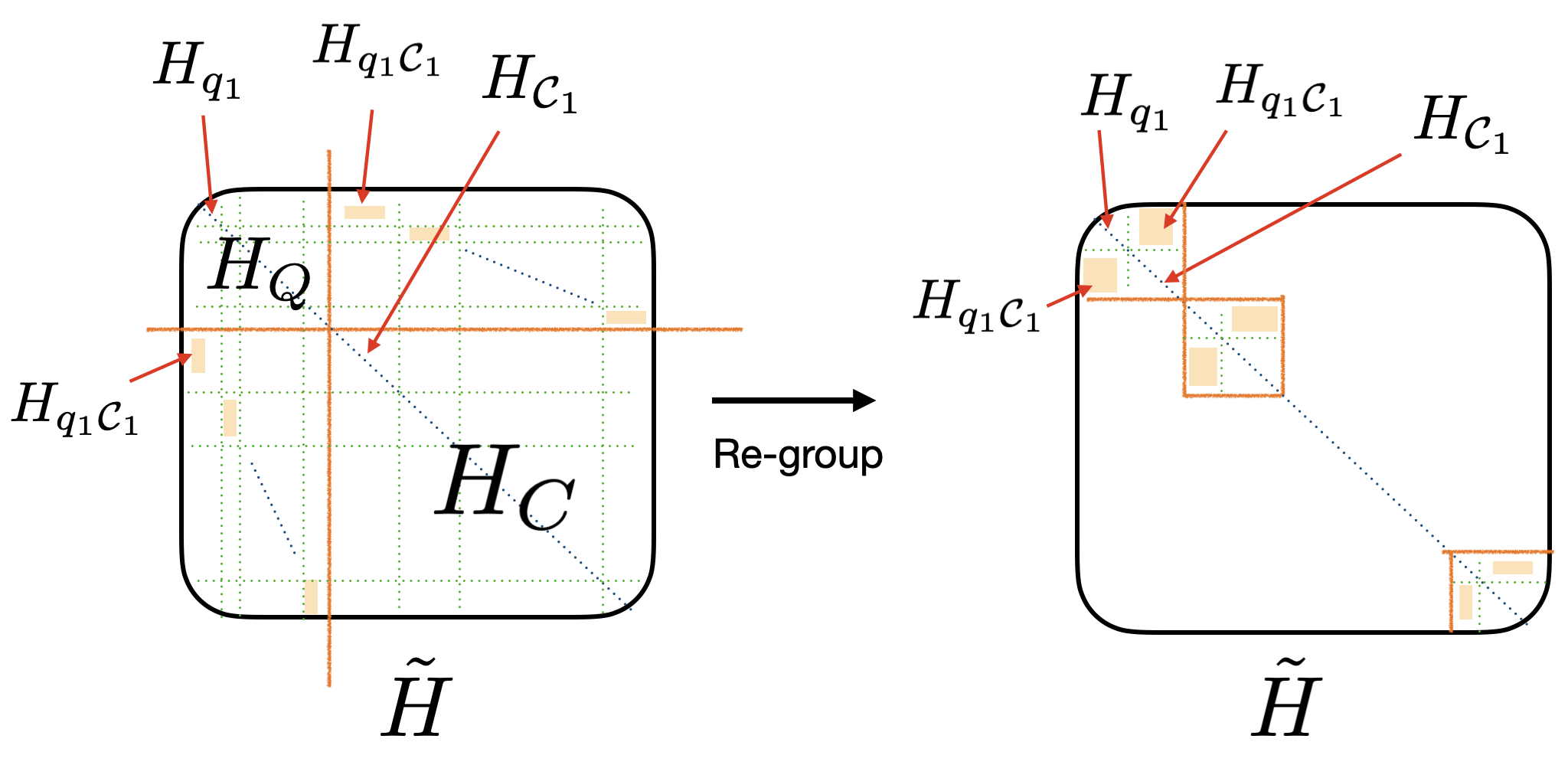}
        \caption{\textbf{Localized decoupling at the Hamiltonian level.} After an appropriate reordering of basis states, the full Hamiltonian becomes block-diagonal, with each block acting on a single qubit and its associated coupler modes. Consequently, each qubit evolves within its own invariant subspace (block).}
        \label{fig:regroup H}
\end{figure}

The two decompositions in Eqs. (\ref{eq.HQC}) and (\ref{eq.Hc}) help to decompose the full Hamiltonian into the qubit-coupler blocks 
\begin{equation}
    \tilde{H}=\sum_{i=1}^N H_i(\lambda),
    \label{eq. H_units}
\end{equation}  
with $H_i(\lambda)=H_{q_i}+H_{Q_i\mathcal{C}i}(\lambda)+H_{\mathcal{C}_i}(\lambda)$.  For notational simplicity, we have removed the tilde symbol throughout these decompositions. This construction partitions the system into $N$ mutually disjoint, non-interacting subspaces, each consisting of a single qubit together with its associated coupler modes. The eigenstates are consequently renormalized by the parameters defined within their respective subgroup elements. In other words, a qubit that is in a subgroup with the coupler modes $c_1$ and $c_2$ acquires a renormalized frequency which is determined solely by these two modes; other qubits and modes do not influence this shift. This makes it clear that we are invoking a physical regrouping into genuinely disjoint subunits. The qubit eigenstates are obviously localized in each subspace, thereby achieving localized decoupling. The process is illustrated in FIG.~\ref{fig:regroup H}.

\subsection{Overlap Method}\label{sec:overlap method}

In order to evaluate effective couplings between two states (qubit-qubit, qubit-readout, etc.) in the non-perturbative regime, we propose a numerical method based on matching eigenvectors and wavefunction delocalization~\cite{Heunisch_2023}. This method is more consistent with our goal to minimize delocalization compared to standard methods such as the least action method~\cite{PhysRevA.101.052308}. 

The common way to calculate effective couplings between two qubits usually employs a transformation to rotate away higher levels and coupler degrees of freedom. An effective Hamiltonian only consisting of the two qubits of interest can be derived from the transformation. The effective couplings can then be read out from the effective Hamiltonian. 

The effective Hamiltonian approach, although commonly adopted by the community, has some flaws. One of the issues is that the effective couplings derived from this approach are frame-dependent. The transformation rotates the original Hamiltonian into a different frame and the effective couplings are not between the original two qubits but the two qubits in the dressed basis in the new frame. As one can choose different transformations to derive the effective Hamiltonian, the effective couplings derived from those transformations will be different. There is no consensus on how to choose the transformation or the frame. Importantly, when we consider delocalization, the choice of transformation/frame is crucial. 

Localization in circuit QED is defined with respect to a local frame, where flux and charge variables are associated with a single node or branch, rather than superpositions across multiple locations. Delocalization and localization are thus measured in this local frame. However, the transformation used to derive the effective Hamiltonian—and in particular to extract the coupling—may redefine the frame in a nonlocal way, thereby introducing delocalization. As a result, decoupling and delocalization may occur simultaneously, since coupling strength and localization are defined in different frames. One might argue that, within the perturbative regime, different choices of transformation should yield equivalent results up to higher-order corrections. 

Including higher orders does not resolve the delocalization, as it is a property of the eigenstates within the fixed local frame. To see this clearly, let us consider a typical two-qubit system with a tunable coupler in between with bare states labeled as $|Q_aCQ_b\rangle$. In the one-excitation subspace, the dressed state of qubit a and b can be expressed in the bare local basis as $\widetilde{|100\rangle}=u_{aa}|100\rangle+u_{ac}|010\rangle+u_{ab}|001\rangle$ and $\widetilde{|001\rangle}=u_{ba}|100\rangle+u_{bc}|010\rangle+u_{bb}|001\rangle$. If the two dressed qubit states are localized, we have $u_{ab}=u_{ba}=0$ and $\widetilde{|100\rangle}=u_{aa}|100\rangle+u_{ac}|010\rangle$, $\widetilde{|001\rangle}=u_{bc}|010\rangle+u_{bb}|001\rangle$. For these states to remain orthogonal, their inner product $\widetilde{\langle 100}\widetilde{|001\rangle}=u_{ac}^*u_{bc}$ must vanish, meaning either $u_{ac}$ or $u_{bc}$ needs to be zero. Without loss of generality, let us assume  $u_{ac}=0$. This means $\widetilde{|100\rangle}=|100\rangle$. The dressed state has to be exactly the same as the bare state. This is impossible in a coupled system, because the qubits interact via the coupler. This contradiction proves that perfect localization is impossible with a single-mode coupler, and any dressed state necessarily has a small but unavoidable component on the other qubit.
Consequently, even when higher-order corrections are included to improve accuracy, the intrinsic delocalization cannot be eliminated. When targeting very high gate fidelities (99.9\%, 99.99\%, or beyond), these unavoidable overlaps can become a non-negligible source of error.

To address the ambiguity in choosing the frame, we propose our localization-compatible method for estimating the qubit-qubit coupling at the operational level. Consider a circuit of qubits and couplers in which $N$ qubits are coupled via $M$ coupling modes. Such a system contains a set of qubit-coupler-mixed eigenstates, which represent $N$ dressed eigenstates for the qubits, $\{\ket{\psi_1},\cdots,\ket{\psi_N}\}$. In this case  the dressed Hamiltonian is diagonal in this basis and denoted by $\tilde{H}_D$ with $\tilde{}$ denoting the interaction being in place and $_{D}$ denoting the diagonal basis, i.e. $\tilde{H}_D \ket{\psi_i}=\tilde{E}_i \ket{\psi_i}$ where $\tilde{E}_i$ denotes the corresponding energy of $\ket{\psi_i}$. 

Moreover, we eliminate all coupling modes from the circuit, qubits are considered  non-interacting and the set of non-interacting states are called the qubit bare states $\{\ket{\varphi_1},\cdots,\ket{\varphi_N}\}$.

Once the dressed eigenbasis has been identified, it is important to recognize that information about interaction strengths is encoded not only in the eigenvalues $\{ \tilde{E}_i \}$, but also in the structure of the associated dressed eigenstates $\{ |\psi_i\rangle \}$. These eigenstates reflect how bare (localized) states are hybridized through interactions, and hence encode both delocalization patterns and effective coupling strengths.

\subsubsection{Validity Domain}

To proceed, we aim to define a well-posed one-to-one mapping between the bare states $\{|\varphi_i\rangle\}$ and dressed states $\{|\psi_i\rangle\}$, under the key assumption that hybridization does not significantly mix multiple bare states. In other words, we assume that each dressed state is dominantly supported on a single bare state, such that:
\begin{equation}
|\langle \varphi_i | \psi_j \rangle|^2 \gg |\langle \varphi_k | \psi_j \rangle|^2, \quad \forall k \neq i.
\end{equation}
This assumption ensures that the overlaps form a near-permutation matrix. More formally, we define the mapping via:
\begin{equation}\label{eq:overlap}
|\psi_i\rangle \equiv \arg \max_{|\psi\rangle} |\langle \varphi_i | \psi \rangle|, \quad |\psi\rangle \in \{ \text{eigenstates of } \tilde{H}_D \},
\end{equation}
with the constraint that this maximum is unique and non-degenerate. That is, we require:
\begin{equation}
\exists! \ |\psi_i\rangle \text{ such that } |\langle \varphi_i | \psi_i \rangle|^2 = \max_{|\psi\rangle} |\langle \varphi_i | \psi \rangle|^2 > \epsilon,
\end{equation}
for some threshold $\epsilon \in (0.5, 1)$ to ensure the dominance of the overlap. This threshold defines the `validity domain' of the method.

If the condition above is not satisfied—i.e., if a dressed state exhibits near-equal support on multiple bare states—the mapping becomes ill-defined. This typically occurs in the strong coupling regime or near accidental degeneracies, where the dressed eigenstates take the form:
\begin{equation}
|\psi_i\rangle \approx \sum_{j=1}^m c_{ij} |\varphi_j\rangle, \quad \text{with } |c_{ij}|^2 \approx \frac{1}{m},
\end{equation}
with $m$ being the number of significantly mixed bare states. In this case, the system cannot be interpreted as weakly dressed versions of localized qubits, and the overlap method loses interpretability, and the states should be assigned manually.

\begin{figure*}[t]
     \centering
     \includegraphics[width=0.9\textwidth]{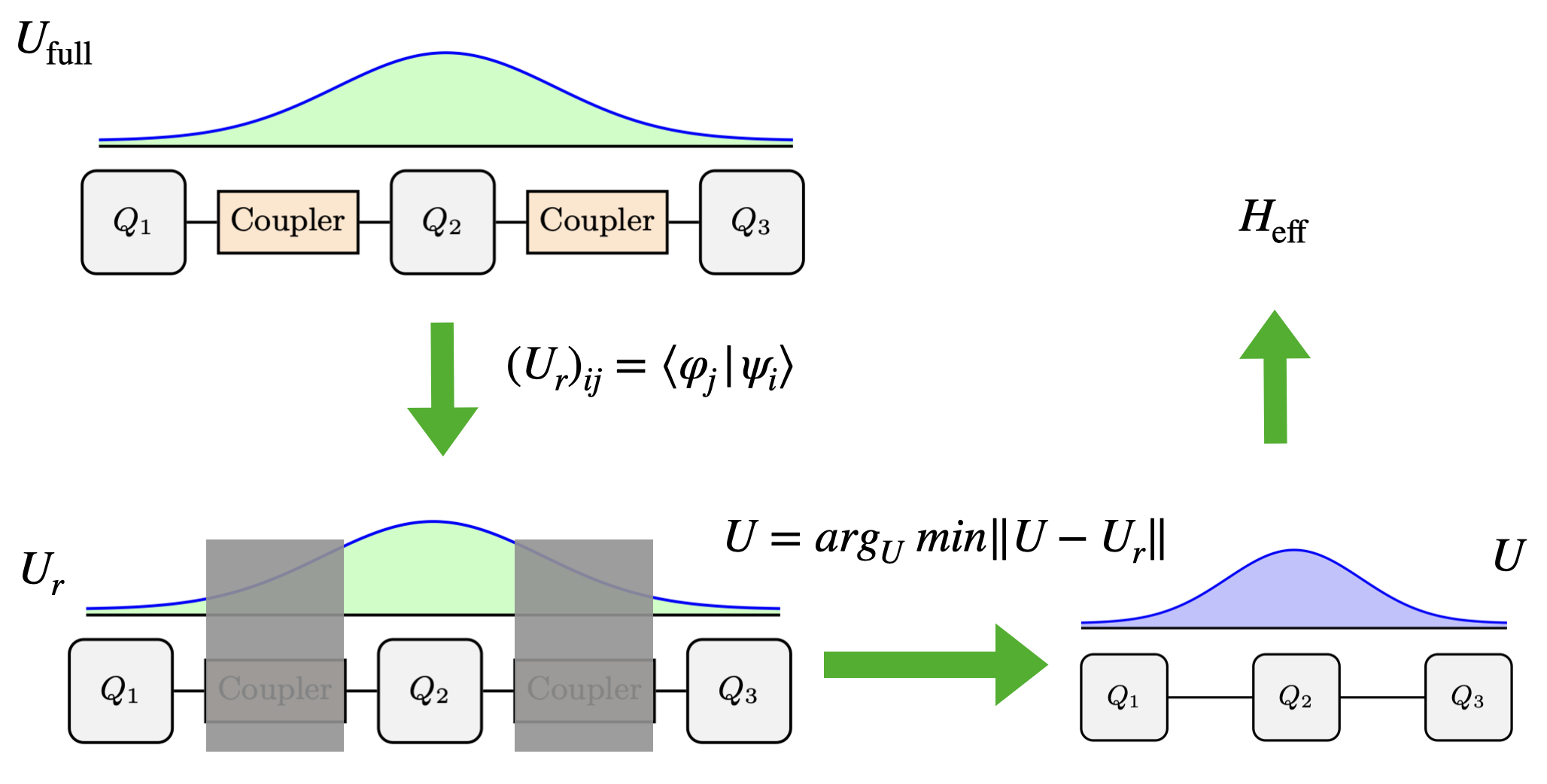}
     \caption{\textbf{Overlap-based construction of effective couplings.} Dressed eigenstates are projected onto a bare subspace to form $U_r$, which is optimally unitarized to obtain $U$. The effective Hamiltonian $H_{\text{eff}}$ is reconstructed from $U$ and dressed energies.}
     \label{fig:overlap method}
\end{figure*}

Therefore, the applicability of this approach is restricted to systems satisfying the following conditions:
\begin{enumerate}
    \item {Spectral separation:} The bare states $\{|\varphi_i\rangle\}$ must be spectrally well-separated such that mixing due to near-degenerate perturbations is suppressed.
    \item {Weak-to-intermediate hybridization:} The interaction strength $J$, in general,  should obey $J \ll \Delta$, where $\Delta$ is the minimum energy gap between target states and their nearest neighbors in the spectrum. The exact threshold, however, depends on the choice of $\epsilon$. When $\epsilon$ is chosen close to $0.5$, the condition can be relaxed to $J \sim \Delta$.
\end{enumerate}

\subsubsection{Overlap-Based Construction of the Effective Hamiltonian}\label{sec:overlap_eff_H}

As discussed above, the method begins with the identification of a reduced subspace $\mathcal{H}_m = \text{span}\{|\varphi_1\rangle, \ldots, |\varphi_m\rangle\}$ consisting of $m$ localized bare states of interest. The full system Hamiltonian $H$ is numerically diagonalized, yielding a set of dressed eigenstates $\{|\psi_i\rangle\}$ and in the previous section we discussed how to associate dressed eigenstates with bare states by a one-to-one mapping, which is physically meaningful under the assumption that coupling is not strong enough to induce widespread hybridization across multiple bare states. When this condition holds, the overlaps $\langle \varphi_i | \psi_j \rangle$ are sharply peaked, and a unique assignment is possible.

Next, we define the reduced diagonalization matrix $U_r$ by projecting the dressed states $\{|\psi_i\rangle\}$ onto the bare subspace:
\begin{equation}
(U_r)_{ij} = \langle \varphi_j | \psi_i \rangle, \quad i,j \in \{1, \ldots, m\}.
\end{equation}
Since $U_r$ is generally not unitary—due to leakage from the subspace or truncation effects—we solve an orthogonal Procrustes problem to find the unitary matrix $U$ that optimally approximates $U_r$ in Frobenius norm \cite{PhysRevA.97.012327}:
\begin{equation}\label{eq:procrustes}
U = \arg\min_{U \in \mathcal{U}(m)} \| U_r - U \|_F.
\end{equation}
This has a well-known closed-form solution: letting $U_r = W \Sigma V^\dagger$ be the singular value decomposition of $U_r$, the minimizing unitary is
\begin{equation}
U = W V^\dagger.
\end{equation}

The effective Hamiltonian within the reduced subspace is then constructed by rotating the diagonal matrix of dressed energies into the bare basis:

\begin{equation}
H_{\text{eff}} = U \, \mathrm{diag}(\tilde{E}_1, \ldots, \tilde{E}_m) \, U^\dagger.
\end{equation}
The off-diagonal elements of $H_{\text{eff}}$ yield effective coupling strengths between bare states:

\begin{equation}\label{eq:J_eff_final}
J^{\text{eff}}_{ij} = \langle \varphi_j | H_{\text{eff}} | \varphi_i \rangle.
\end{equation}

\medskip

\subsubsection{Block-wise Application and Spatial Locality}

The method naturally generalizes to large systems by applying it block-wise over localized regions. For example, in circuit QED architectures with modular qubit-coupler units, the full Hamiltonian can be approximately decomposed into quasi-independent blocks:

\begin{equation}
H_{\text{full}} \approx \bigoplus_{\text{blocks}} H_{\text{eff}}^{(\text{block})},
\label{eq:blocks}
\end{equation}
where each $H_{\text{eff}}^{(\text{block})}$ is constructed independently via the overlap method in a local subspace. This modular approach significantly reduces computational complexity and respects the underlying connectivity of the hardware.

\begin{figure*}[t]
     \centering
     \includegraphics[width=0.9\textwidth]{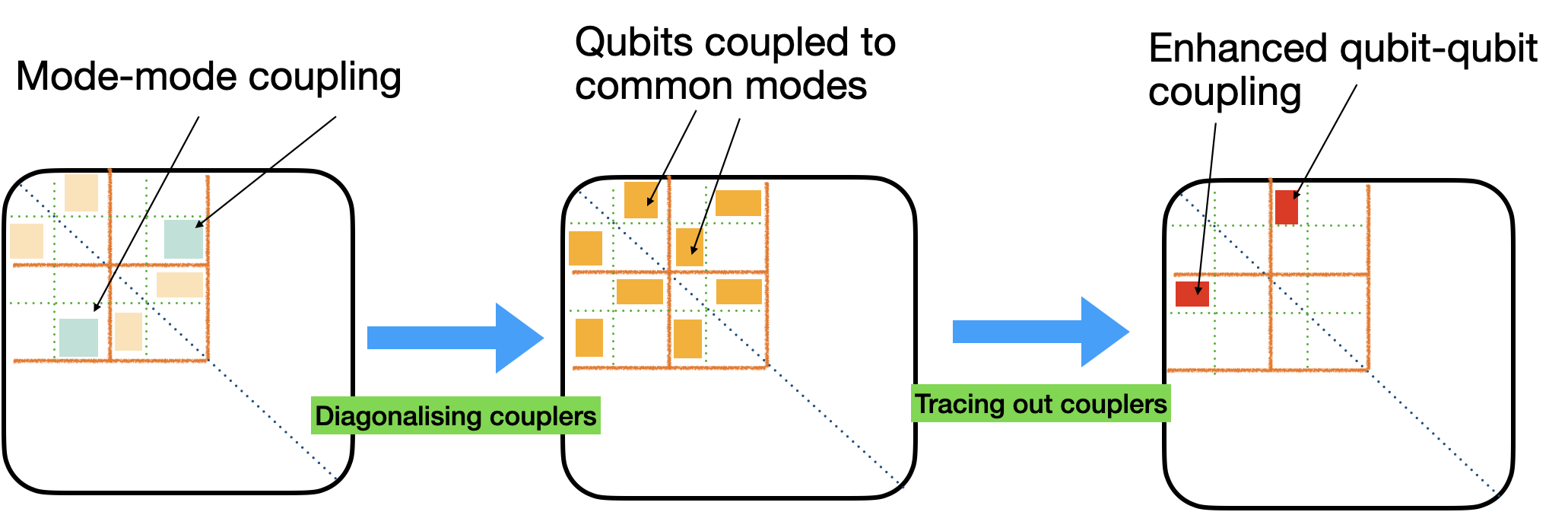}
     \caption{\textbf{Enhanced qubit-qubit coupling via mode sharing.} Mode-mode interactions hybridize the coupler spectrum, allowing both qubits to couple to shared modes. Tuning $\lambda$ controls the interaction strength.}
     \label{fig: enhance qq}
\end{figure*}

The resulting effective Hamiltonian satisfies several key properties:

\begin{enumerate}
    \item {Spectral Consistency:} The eigenvalue differences within the effective Hamiltonian preserve those from the full system:
    \[
    \tilde{E}_i - \tilde{E}_j = E_i^{\text{eff}} - E_j^{\text{eff}}.
    \]
    \item {Minimal Basis Distortion:} The transformation $U$ is the closest unitary to the bare-dressed rotation matrix $U_r$, minimizing frame mismatch and preserving localization structure.
    \item Delocalization Sensitivity: In the case where each dressed state is fully localized (i.e., $|\psi_i\rangle = |\varphi_i\rangle$), the overlap matrix becomes identity, and the resulting coupling vanishes: $J^{\text{eff}}_{ij} = 0$ for $i \neq j$.
\end{enumerate}

\subsection{Modifying Qubit-Qubit Coupling via Mode-Mode Interaction $\lambda$}\label{sec:enhance_J}

We now examine whether the same tunable parameter $\lambda$, previously used to decouple idle qubits, can be repurposed to enhance qubit-qubit interactions during gate operation.

Consider a system composed of multiple qubits and associated coupler modes. Initially, the Hamiltonian factorizes into decoupled components:
\begin{equation}
H = \sum_i H_i, \quad [H_i, H_j] = 0,
\end{equation}
where $H_i$ includes the Hamiltonian of qubit $Q_i$ and its local coupler modes.

To activate coupling between two qubits (e.g., $Q_1$ and $Q_2$), we introduce mode-mode interactions between their respective coupler modes:
\begin{equation}
H = H_{12} + \sum_{k \geq 3} H_k,
\end{equation}
with
\begin{equation}
H_{12} = H_{q_1} + H_{q_2} + \sum_{i,j \in \mathcal{C}_1 \cup \mathcal{C}_2} H_{c_i c_j}(\lambda),
\end{equation}
where $\mathcal{C}_1$ and $\mathcal{C}_2$ denote the sets of coupler modes attached to $Q_1$ and $Q_2$, respectively, and $H_{c_k c_l}(\lambda)$ encodes the mode-mode interactions tunable via $\lambda$.
This configuration forms a local interacting block comprising $Q_1$, $Q_2$, and their associated coupler modes. The rest of the system remains dynamically decoupled.

Expanding $H_{12}$, we separate direct and indirect interaction terms:
\begin{equation}
\begin{split}
H_{12} = &\sum_{i=1}^2 \left( H_{q_i} + H_{Q_i, \mathcal{C}_i} \right) \\
&+ \sum_{i,j \in \mathcal{C}_1 \cup \mathcal{C}_2} H_{c_i c_j}(\lambda),
\end{split}
\end{equation}
where $H_{Q_i, \mathcal{C}_i}$ is the coupling between $Q_i$ and its local coupler modes $\mathcal{C}_i$.

We now diagonalize the coupled-mode Hamiltonian using a unitary transformation $U_c$ such that:
\begin{equation}
    U_c^\dagger(\lambda) \left( \sum_{i,j} H_{c_i c_j}(\lambda) \right) U_c(\lambda) = \sum_k H_{c_k}^{\text{diag}}.
\end{equation}

In this new frame, the coupler modes are de-hybridized, therefore when the transformation is applied on the mode-qubit interaction, it results in modified interaction strength with the orthonormal modes:
\begin{equation}
\begin{split}
    \tilde{H}_{12}=&U_c^\dagger(\lambda) H_{12}U_c(\lambda)\\
    =&\sum_{i=1}^2 H_{q_i} + \sum_k \left( H_{q_1 c_k}(\lambda) + H_{q_2 c_k}(\lambda) + H_{c_k}^{\text{diag}} \right)
\end{split}
\end{equation}
where $k$ runs over the diagonalized hybrid modes. 

This frame reveals that both qubits are coupled to shared coupler modes, enabling indirect qubit-qubit interaction. The strength of this interaction depends sensitively on $\lambda$ through the structure of the hybrid modes and their coupling amplitudes.

To quantify this enhancement, we perform a simple estimation via the Schrieffer–Wolff Transformation.
Consider a simplified model where $Q_1$ and $Q_2$ couple to $m$ identical harmonic modes with strength $g_0$. In the diagonal frame, the Hamiltonian reads~\cite{RevModPhys.93.025005}:
\begin{equation}
\begin{split}
\tilde{H}_{12} &= -\frac{\omega_{q1}}{2} \sigma_1^z - \frac{\omega_{q2}}{2} \sigma_2^z + \sum_{k=1}^m \omega_{k} c_k^\dagger c_k \\
&\quad + g_0 \sum_{k=1}^m \left( \sigma_1^x + \sigma_2^x \right) \otimes (c_k + c_k^\dagger),
\end{split}
\end{equation}
where $\sigma^x = \sigma^+ + \sigma^-$ denotes transverse coupling.

Applying a second-order Schrieffer–Wolff transformation to eliminate the coupler degrees of freedom and the rotating-wave approximation  yields the effective two-qubit Hamiltonian:
\begin{equation}
H_{\text{eff}} = -\frac{\omega_{q1}}{2} \sigma_1^z - \frac{\omega_{q2}}{2} \sigma_2^z + J \left( \sigma_1^+ \sigma_2^- + \sigma_1^- \sigma_2^+ \right),
\end{equation}
with the exchange coupling:
\begin{equation}
\begin{split}
J =-\frac{g_0^2}{2} \sum_{k=1}^m  \Bigg( &
    \frac{1}{\omega_k - \omega_{q1}}
  + \frac{1}{\omega_k + \omega_{q1}}\\
  &+ \frac{1}{\omega_k - \omega_{q2}}
  + \frac{1}{\omega_k + \omega_{q2}}
\Bigg).
\end{split}
\end{equation}

Assuming similar detunings $\omega_{q1} \approx \omega_{q2}$ and approximately degenerate coupler modes $\omega_{k} \approx \omega_c$, this simplifies to: 

\begin{equation}
    J \approx m \cdot \frac{g_0^2}{\Delta}, \quad \Delta = \omega_q - \omega_c.
\end{equation}

Thus, the effective coupling strength scales linearly with the number of hybrid modes $m$, showing that interaction strength can be amplified by mode sharing—a mechanism tunable via $\lambda$.

This analysis highlights a dual utility of the parameter $\lambda$: it can be tuned to minimize residual couplings during idle, and also to boost inter-qubit coupling on demand during gate operation by activating constructive mode-mediated interactions.

\section{Full Localization: Ideal Decoupling}\label{sec:full_localization}

We demonstrate how to achieve complete decoupling between two transmons by controlling the interaction strength between a two-mode coupler and following the general recipe outlined in Sec.~\ref{sec:general theo}. A circuit-level derivation starting from the Lagrangian is sketched in Appendix~\ref{appen:Lag}.

We consider a circuit with two qubits $Q_a$ and $Q_b$, where they are coupled to a two-mode coupler with modes $c_1$ and $c_2$. The coupler Hamiltonian part of the system $H_C=\sum_{i=1,2}\omega_{i}c_i^\dagger c_i-\lambda(c_1-c_1^\dagger )(c_2-c_2^\dagger)$   can be fully diagonalized (after applying the rotating wave approximation eliminating doubly excited/de-excited states) via a Bogoliubov transformation \cite{RevModPhys.93.025005,Boissonneault_2008,622206}:
\begin{equation}\label{eq:Bogoliubov}
    U_C = e^{\Lambda (c_1^\dagger c_2 - c_1 c_2^\dag)}.
\end{equation}
This transforms the mode operators as:
\begin{equation}
\begin{split}
U_C^\dagger c_1 U_C &= \cos\Lambda \, c_1 + \sin\Lambda \, c_2, \\
U_C^\dagger c_2 U_C &= \cos\Lambda \, c_2 - \sin\Lambda \, c_1.
\end{split}
\label{eq.bogoli}
\end{equation}

Choosing the mixing angle $\Lambda = \frac{1}{2} \arctan(2\lambda/\Delta_{12})$, with $\Delta_{12} = \omega_2 - \omega_1$, diagonalizes the coupler Hamiltonian:
\begin{equation}
U_C^\dagger H_C U_C = \tilde{\omega}_1 c_1^\dagger c_1 + \tilde{\omega}_2 c_2^\dagger c_2,
\end{equation}
with
\begin{align}
\tilde{\omega}_{1,2} = \frac{1}{2}\left(\omega_1 + \omega_2 \mp \sqrt{\Delta_{12}^2 + 4 \lambda^2} \right).
\end{align}

Under this transformation, the qubit-coupler interaction Hamiltonian $\sum_{i=a,b}\sum_{k=1,2} g_{ik} \left(q_i^\dagger c_k+H.c.\right)$ becomes:
\begin{equation}\label{eq:BogoliubovHint}
\begin{split}
\tilde{H}_{\text{int}} = U_C^\dagger H_{\text{int}} U_C &= \sum_{i=1}^2 \sum_{k=1}^2 \tilde{g}_{ik} \left(q_i^\dagger c_k + q_i c_k^\dagger\right),
\end{split}
\end{equation}
where $q_1 = a$, $q_2 = b$ are the annihilation operators of the two transmons, and the transformed couplings $\tilde{g}_{ij}$ are given by:
\begin{equation}
\begin{pmatrix}
\tilde{g}_{a1} \\ \tilde{g}_{a2}
\end{pmatrix}
= R(\Lambda)
\begin{pmatrix}
g_{a1} \\ g_{a2}
\end{pmatrix}, \quad
\begin{pmatrix}
\tilde{g}_{b1} \\ \tilde{g}_{b2}
\end{pmatrix}
= R(\Lambda)
\begin{pmatrix}
g_{b1} \\ g_{b2}
\end{pmatrix},
\end{equation}
with the rotation matrix:
\begin{equation}
R(\Lambda) = 
\begin{pmatrix}
\cos\Lambda & -\sin\Lambda \\
\sin\Lambda & \cos\Lambda
\end{pmatrix}.
\end{equation}

\subsection{Condition for Full Decoupling}

To fully decouple the two qubits from one another, we need to impose a condition on the two coupling mode interaction strength. This requires the following condition:
\[
\tilde{g}_{a2} = 0, \quad \text{and}\quad  \tilde{g}_{b1} = 0.
\]
This leads to the condition:
\begin{equation}\label{eq:decoupling_cond}
\tan\Lambda = -\frac{g_{a2}}{g_{a1}}, \quad \text{and} \quad \tan\Lambda= \frac{g_{b1}}{g_{b2}}.
\end{equation}

Using $\Lambda = \frac{1}{2} \arctan(2\lambda/\Delta_{12})$, and assuming symmetric coupling ratios $g_{a2} = -\alpha g_{a1}$ and $g_{b1} = \alpha g_{b2}$, the condition simplifies to: 
\begin{equation}
\frac{\lambda}{\Delta_{12}} = \frac{\alpha}{1 - \alpha^2}.
\label{eq. decoupling2}
\end{equation}

\subsection{Resulting Decoupled Hamiltonian}

With this condition, the interaction Hamiltonian reduces to:
\begin{equation}
\tilde{H}_{\text{int}} = \frac{1}{\cos\Lambda} \left[ g_{a1}(a^\dagger c_1 + a c_1^\dag) + g_{b2}(b^\dagger c_2 + b c_2^\dag) \right],
\end{equation}
and the full system separates into two uncoupled subsystems:
\begin{widetext}
\begin{equation}
\begin{split}
\tilde{H} = U_C^\dagger H U_C &= \omega_a a^\dagger a + \frac{\delta_a}{2}(a^\dagger a - 1)a^\dagger a + \tilde{\omega}_1 c_1^\dagger c_1 + \frac{g_{a1}}{\cos\Lambda}(a^\dagger c_1 + a c_1^\dag) \\
&+ \omega_b b^\dagger b + \frac{\delta_b}{2}(b^\dagger b - 1)b^\dagger b + \tilde{\omega}_2 c_2^\dagger c_2 + \frac{g_{b2}}{\cos\Lambda}(b^\dagger c_2 + b c_2^\dag).
\end{split}
\end{equation}
\end{widetext}
Each transmon couples exclusively to a single hybridized mode. The absence of any shared coupling path guarantees strict decoupling between the two qubits, realizing ideal localization in the diagonalized frame.

\begin{figure*}[t]
\centering
\includegraphics[width=0.245\textwidth]{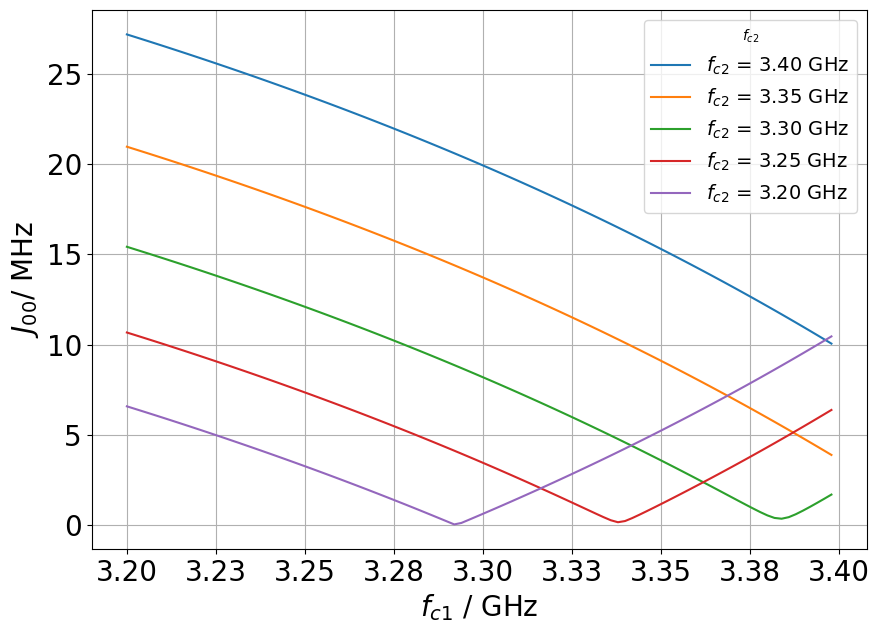}\put(-90,98){\textbf {(a)}}
\includegraphics[width=0.245\textwidth]{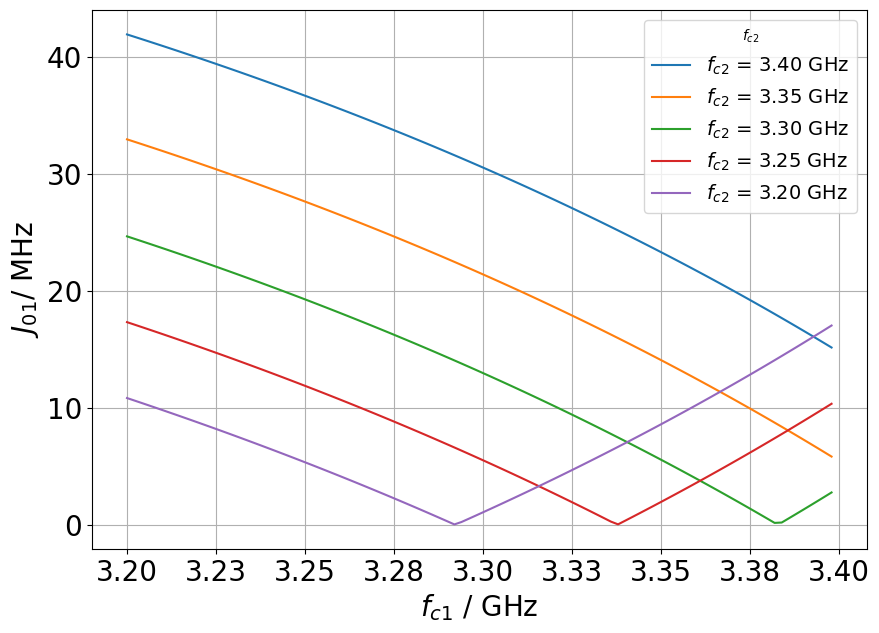}\put(-90,98){\textbf {(b)}}
\includegraphics[width=0.245\textwidth]{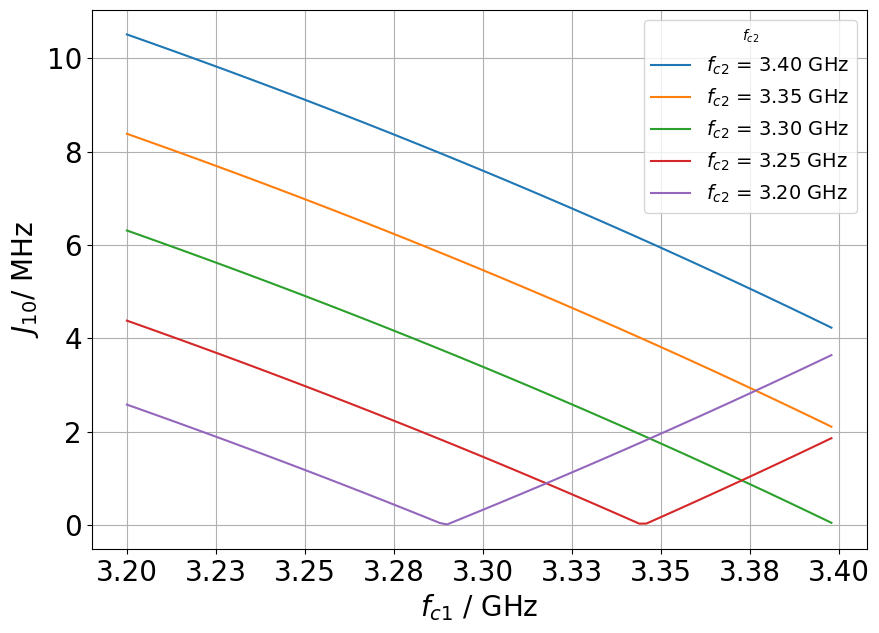}\put(-90,98){\textbf {(c)}}
\includegraphics[width=0.245\textwidth]{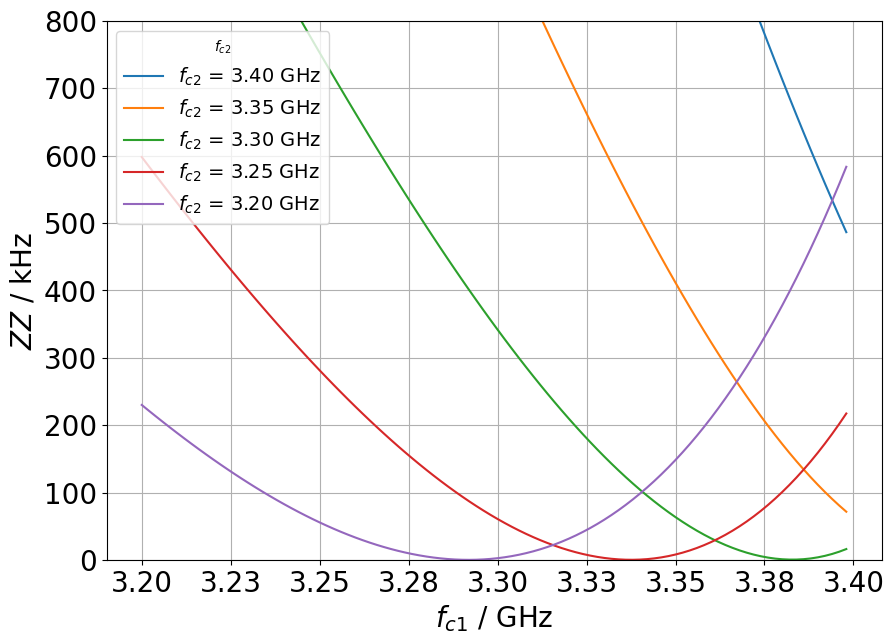}\put(-90,98){\textbf {(d)}}
\vspace{-0.1in}
\caption{\textbf{$J$ couplings and $ZZ$ interaction in the off-mode.} The two transmons are detuned by 600 MHz, corresponding to the idle point. In the simulation, we vary the frequencies of modes 1 and 2. The fully localized decoupling zone is where all $J$'s and $ZZ$ are suppressed to nearly zero. The effective couplings $J$s are calculated using a perturbative expansion of the overlap method. As discussed in Sec.~\ref{sec:overlap method}, with the overlap method, zero $J$ indicates zero delocalization. (a) Interaction in the one-excitation subspace $J_{00}$.  (b)(c) Interaction in the two-excitation subspace $J_{01}$ and $J_{10}$. They show similar trends as $J_{00}$. (d) $ZZ$ interaction. Minimal $ZZ$ is suppressed close to zero.}
\label{fig:Js off}
\end{figure*}

FIG.~\ref{fig:Js off} illustrates a numerical simulation of the effective couplings $J$ within this fully localized regime. The overlap-based extraction method for $J$ is detailed in Appendix~\ref{appen:J}, and the simulation parameters are summarized in Table~\ref{tab:para}. Non-RWA terms are included in the numerical simulation. Interactions are assumed to be of the form: $-(q_i-q_i^\dag)(c_j-c_j^\dag)$ and $-(c_i-c_i^\dag)(c_j-c_j^\dag)$.

Let us label our states in the excitation basis as $\ket{q_a,  c_1,  c_2,  q_b}$, denoting the states of qubit $a$, coupler mode 1, coupler mode 2 and transmon $b$. It should be noted that, when calculating the effective couplings from $|1001\rangle$, although coupler states such as $|1100\rangle, |1010\rangle...$ also participate in interaction, we only include the overlap with transmon states $|2000\rangle$ and $|0002\rangle$ in the final formula of $J^\text{eff}$. This is because we only want to ensure zero delocalization among qubits but not between qubits and the coupler. The interactions mediated by the coupler are already taken into account when we numerically diagonalize the full Hamiltonian.

\begin{table}[h]
    \centering
    \begin{tabular}{|c|c|}
        \hline
        Parameter & Value \\
        \hline
        $\omega_a/2\pi$ & 4.2 GHz \\
        \hline
        $\omega_b/2\pi$ & 3.6 GHz \\
        \hline
        $\delta_a/2\pi$ & -300 MHz \\
        \hline
        $\delta_b/2\pi$ & -350 MHz \\
        \hline
        $g_{ab}/2\pi$ & 4 MHz \\
        \hline
        $g_{a1}/2\pi$ & 150 MHz \\
        \hline
        $g_{a2}/2\pi$ & -200 MHz \\
        \hline
        $g_{b1}/2\pi$ & 150 MHz \\
        \hline
        $g_{b2}/2\pi$ & 150 MHz \\
        \hline
        $g_{12}/2\pi$ & 10 MHz \\
        \hline
    \end{tabular}
    \caption{Parameters used in the numerical simulation. The negative qubit-coupler coupling can be realized by designing the parities of the coupler modes.}
    \label{tab:para}
\end{table}

In FIG.~\ref{fig:Js off}, for each $f_{c2}$, all three effective $J$ couplings and $ZZ$ interaction reach minima around similar $f_{c1}$, meaning all interactions can be turned off at the same coupler spot. We find that by optimizing the coupler frequencies, the minimal $J_{00}$ can be suppressed below 10 kHz. The details can be found in Appendix~\ref{appen:mini J}.

\subsection{Multi-mode coupler circuit design}

\begin{figure*}
    \centering
    \includegraphics[width=0.9\linewidth]{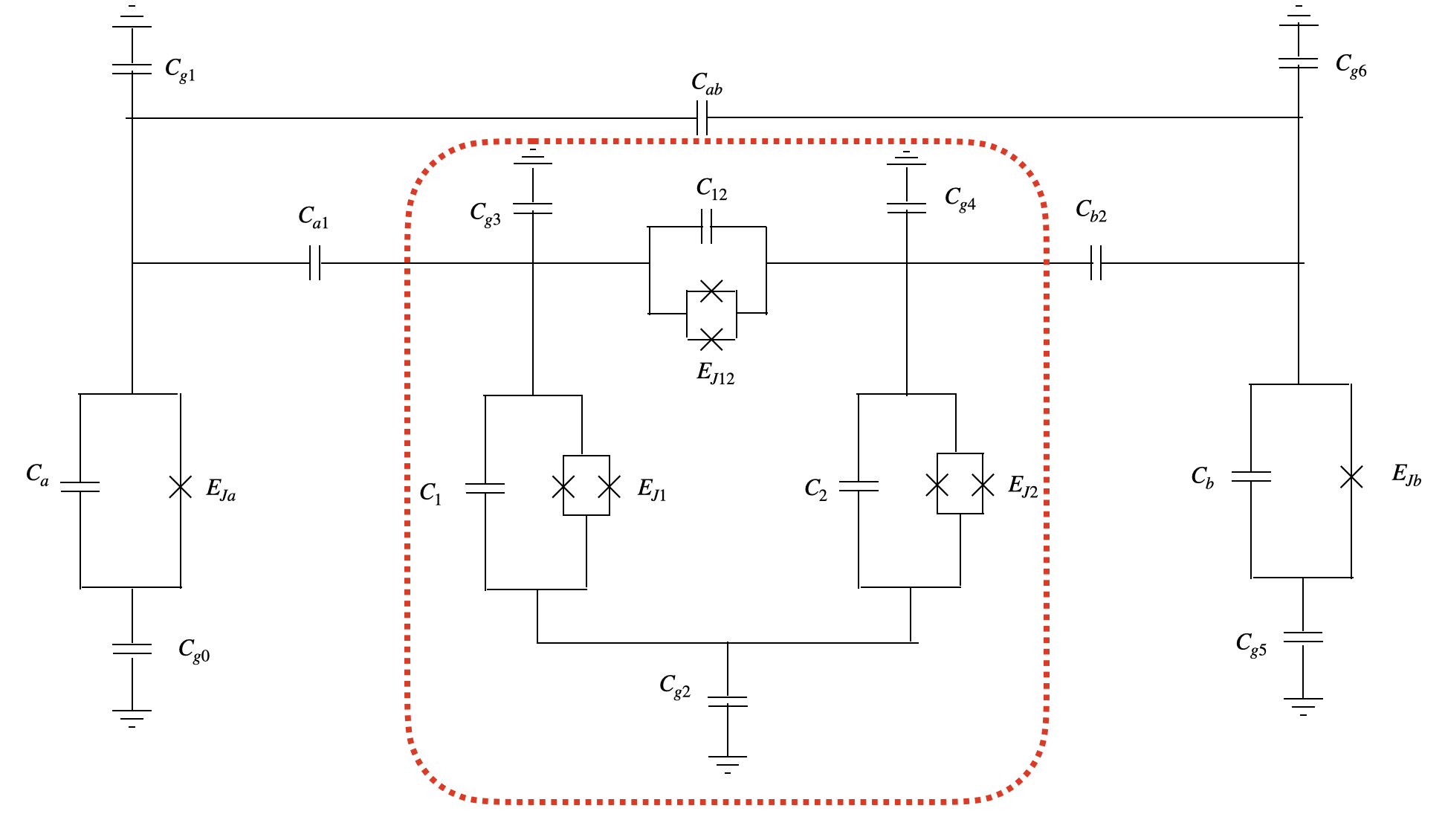}
    \put(-380,250){\textbf {(a)}}\\
    \includegraphics[width=0.9\linewidth]{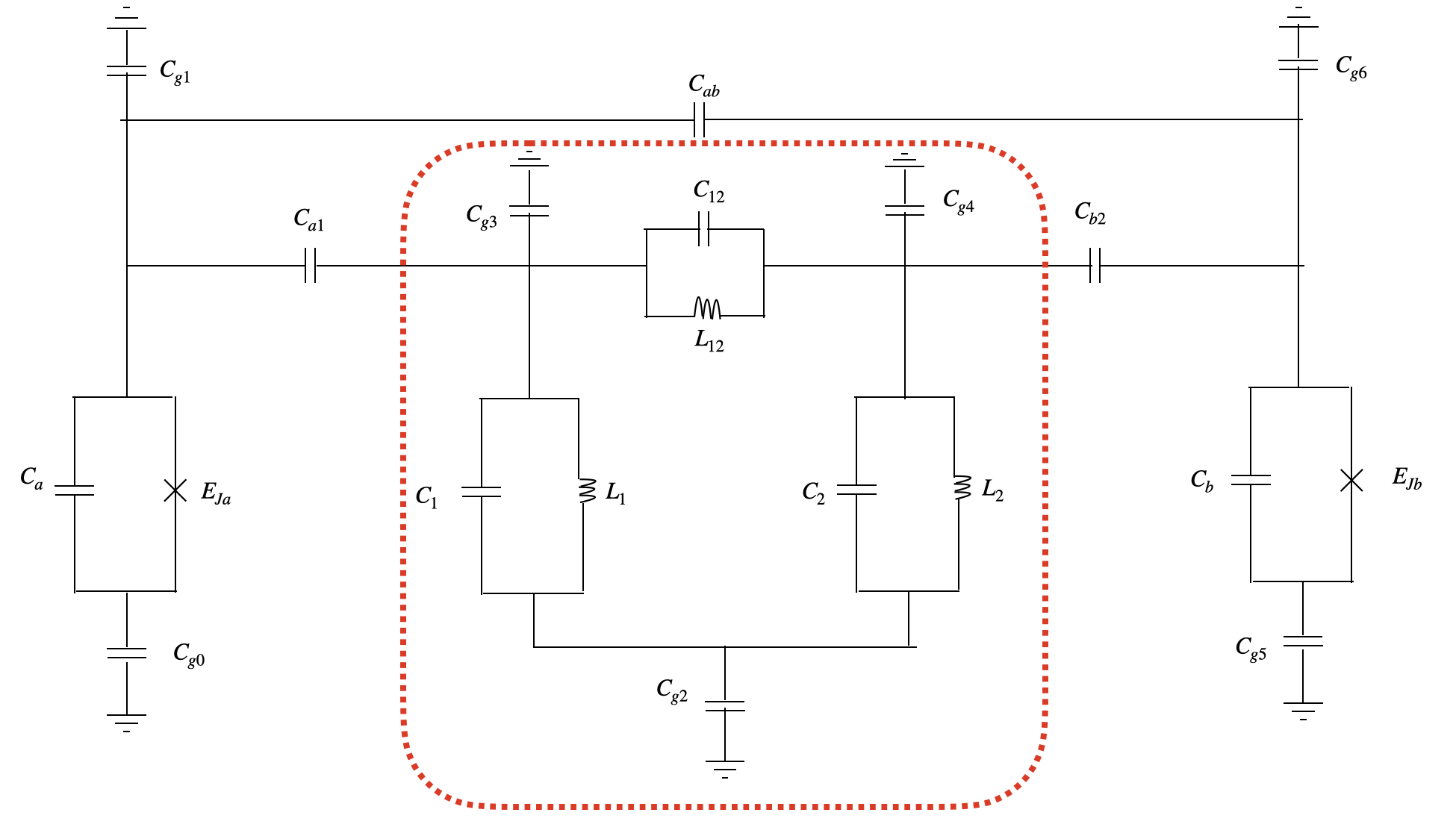}
    \put(-380,250){\textbf {(b)}}
    \caption{\textbf{Schematic diagram of the two proposed coupler circuit designs coupling two fixed-frequency transmons.} Following Ref.~\cite{PhysRevApplied.15.064063}, a floating design is implemented, utilizing grounding capacitors for the qubits and the coupler. The coupler parts are highlighted within the red dashed box (a) The SQUID $\Delta-$network design. The coupler has three SQUIDs as tunable elements. We assume symmetric junctions for each SQUID. The coupler is connected to qubits from the two sides via capacitors $C_{a1}$ and $C_{b2}$. (b) The tunable inductance $\Delta-$network design. The tunable SQUIDs in (a) are replaced by tunable linear inductors labeled as $L_1, L_2$ and $L_{12}$. The tunable inductors can be fabricated as series of SQUIDs}
    \label{fig:circuit designs l and r}
\end{figure*}

We outline two lumped-element circuit designs that satisfy these localization conditions. These circuit designs can hold multiple modes with tunable frequencies and g couplings. Motivated by the metamaterial couplers\cite{Indrajeet_2020,PRXQuantum.5.020325} and the double-transmon coupler\cite{Goto_2022,PhysRevX.14.041050}, we propose two coupler circuit designs. They are the SQUID~(superconducting quantum interference devices) $\Delta-$network design and the tunable inductance $\Delta-$network design shown in  FIG.~\ref{fig:circuit designs l and r}. We employ a floating architecture, incorporating grounding capacitors on both the qubits and the coupler \cite{PhysRevApplied.15.064063}. In the SQUID design, the three SQUIDs serve as tunable elements in the coupler. In fact, our SQUID design is a direct extension of the double-transmon design\cite{Goto_2022,PhysRevX.14.041050}. The primary modification in our approach is the replacement of single Josephson junctions with SQUIDs, providing an additional degree of freedom for in-situ tuning. In the tunable inductance design, the tunable inductors can be implemented in experiments as a series of SQUIDs. The inductance is tunable by applying magnetic flux to the SQUID array. It has been experimentally demonstrated in different works\cite{PhysRevLett.109.137003,Kim_Shrekenhamer_McElroy_Strikwerda_Alldredge_2019}.

The details of the derivation of the circuit Lagrangian and Hamiltonian can be found in Appendix.~\ref{appen:circ}. Here we summarize the main results. After removing redundant variables and assuming symmetric junctions everywhere, the Lagrangian of the full circuit in FIG.~\ref{fig:circuit designs l and r}(a) reads

\begin{equation}\label{eq:3squid L}
\begin{split}
        \mathcal{L}=&\frac{1}{2}\dot{\boldsymbol{\Phi}}^T\mathbf{C}\dot{\boldsymbol{\Phi}}+E_{Ja}\cos(\frac{2\pi}{\Phi_0}\Phi_a)+E_{Jb}\cos(\frac{2\pi}{\Phi_0}\Phi_b)\\
        &+2E_{J1}\cos(\frac{\pi}{\Phi_0}\Phi_{e1})\cos(\frac{2\pi}{\Phi_0}\Phi_1)\\
        &+2E_{J2}\cos(\frac{\pi}{\Phi_0}\Phi_{e2})\cos(\frac{2\pi}{\Phi_0}\Phi_2) \\
        & +2E_{J12}\cos(\frac{\pi}{\Phi_0}\Phi_c)\cos(\frac{2\pi}{\Phi_0}\Phi_1+\frac{2\pi}{\Phi_0}\Phi_2)  
\end{split}
\end{equation}

Defining the conjugate variables $Q_k=\frac{\partial \mathcal{L}}{\partial \dot{\Phi}_k}$ and the reduced variables $\varphi_k=\frac{2\pi}{\Phi_0}\Phi_k, n_k=\frac{Q_k}{2e}$, the Hamiltonian is

\begin{equation}\label{eq:3SQUID H}
\begin{split}
        H=&4\boldsymbol{n}^T\mathbf{E}_C\boldsymbol{n}-E_{Ja}\cos(\varphi_a)-E_{Jb}\cos(\varphi_b)\\
        &-\tilde{E}_{J1}\cos(\varphi_1)-\tilde{E}_{J2}\cos(\varphi_2) \\
        & -\tilde{E}_{J12}\cos(\varphi_1+\varphi_2)  
\end{split}
\end{equation}
where we have defined:  $\mathbf{E}_C=\frac{e^2\mathbf{C}^{-1}}{2},\tilde{E}_{J1}=2E_{J1}\cos(\varphi_{e1}/2),\tilde{E}_{J2}=2E_{J2}\cos(\varphi_{e2}/2)$ and $\tilde{E}_{J12}=2E_{J12}\cos(\varphi_{c}/2)$.
We then introduce the creation and annihilation operators $a^\dag_k, a_k$ with $k=a,b,1,2$ to partially diagonalize the Hamiltonian:
\begin{equation}\label{eq:3SQUID cre anni}
    \begin{split}
        \varphi_k=&\left(\frac{2E_{C,kk}}{E_{Jk}}\right)^{1/4}(a^\dag_k+a_k)\\
        n_k=&\frac{i}{2}\left(\frac{E_{Jk}}{2E_{C,kk}}\right)^{1/4}(a^\dag_k-a_k)
    \end{split}
\end{equation}
    
In Eq.~(\ref{eq:3SQUID H}), the frequencies of the two coupler modes 1 and 2 are tunable by tuning $E_{J1}$ and $E_{J2}$. The mode-mode coupling can be controlled by $E_{J12}$. These Josephson energies can be tuned by applying external magnetic fluxes.
For the numerical simulations, we use the Hamiltonian in Eq.~(\ref{eq:3SQUID H}) and the creation and annihilation operators defined in Eq.~(\ref{eq:3SQUID cre anni}). The parameters are listed in TABLE~\ref{tab:para 3s}.

\begin{table}[h]
    \centering
    \begin{tabular}{|c|c|}
        \hline
        Parameter & Value \\
        \hline
        $E_{Ja}$ & 14 GHz \\
        \hline
        $E_{Jb}$ & 18 GHz \\
        \hline
        $E_{J1}$ & 4 GHz \\
        \hline
        $E_{J2}$ & 4 GHz \\
        \hline
        $E_{J12}$ & 7.5 GHz \\
        \hline
        $C_a$ & 70 fF \\
        \hline
        $C_b$ & 70 fF \\
        \hline
        $C_1$ & 70 fF \\
        \hline
        $C_2$ & 80 fF \\
        \hline
        $C_{ab}$ & 0.05 fF \\
        \hline
        $C_{12}$ & 20 fF \\
        \hline
        $C_{a1}$ & 10 fF \\
        \hline
        $C_{b2}$ & 10 fF \\
        \hline
        $C_{gi}$ & 50 fF \\
        \hline
    \end{tabular}
    \caption{Parameters used in the numerical simulation of the SQUID $\Delta$-network coupler.}
    \label{tab:para 3s}
\end{table}

\begin{figure*}
    \centering
    \includegraphics[width=0.9\textwidth]{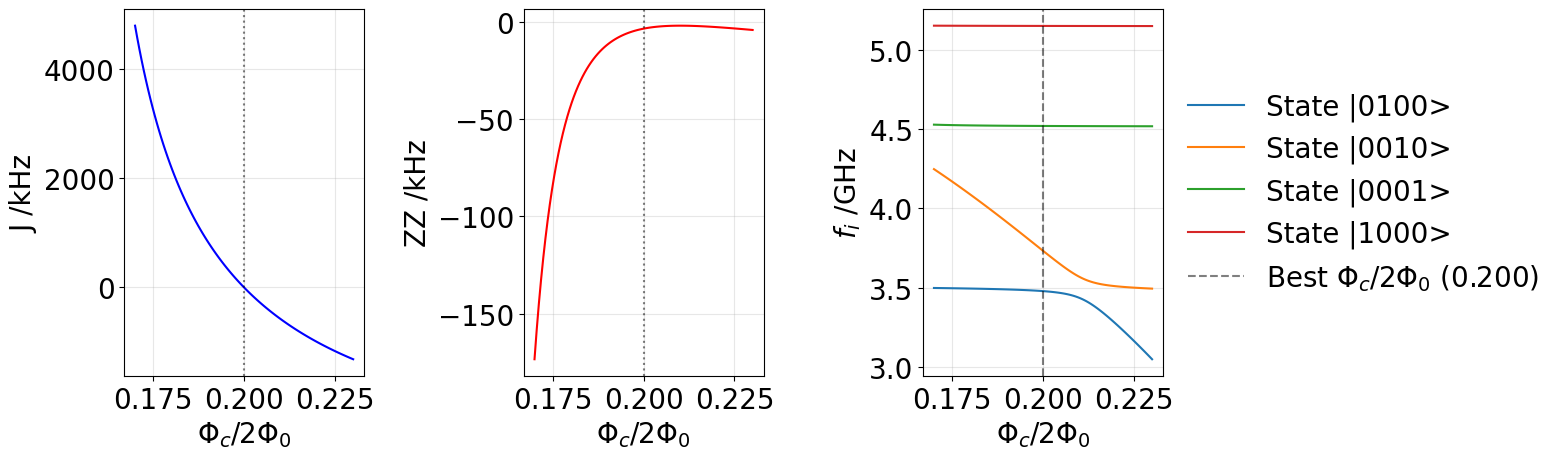}\put(-440,140){\textbf {(a)}} \put(-320,140){\textbf {(b)}}\put(-200,140){\textbf {(c)}}\\
    \includegraphics[width=0.9\textwidth]{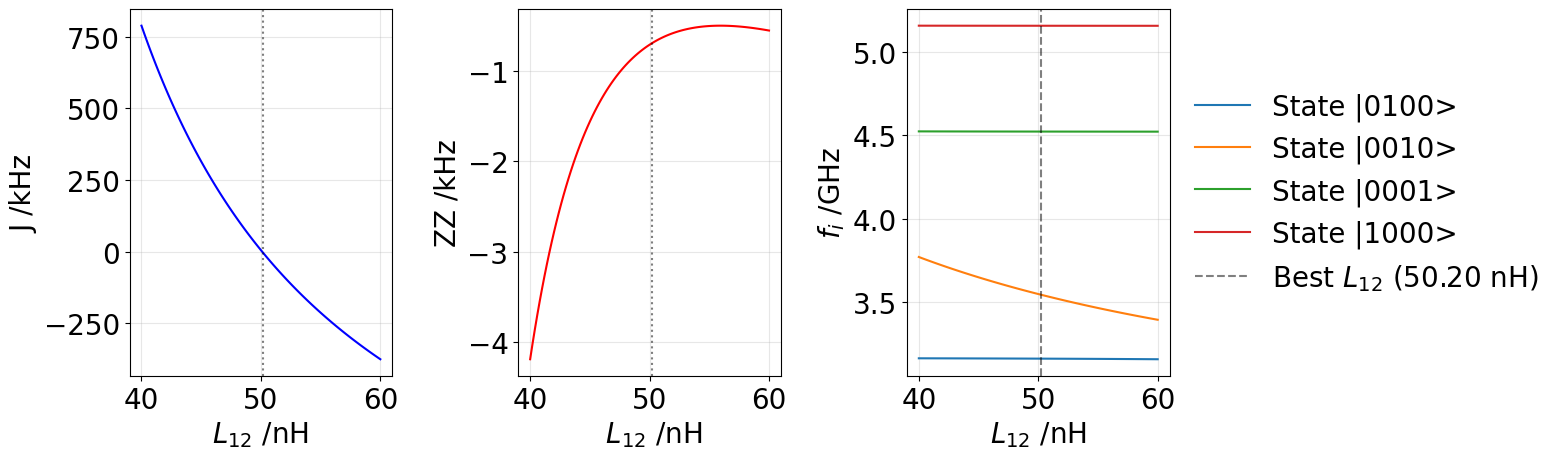}\put(-440,140){\textbf {(d)}} \put(-320,140){\textbf {(e)}}\put(-200,140){\textbf {(f)}}
    \caption{\textbf{Numerical results of the full system with $\Delta$-network couplers.} The effective coupling $J_{00}$, residual $ZZ$ interactions and the spectra are plotted as functions of the tunbale elements. The optimal $J$ and $ZZ$ trad-off points are marked by the dotted lines. (a)Results from the SQUID design. Both $J$ and $ZZ$ can reach zero. At the optimal point, $J=-18.1$ kHz and $ZZ=-3.7$ kHz (b)Results from the tunable inductance design. Both $J$ can reach zero but the zero $ZZ$ point is out of the scan range. At the optimal point, $J=-5.0$ kHz and $ZZ=-0.7$ kHz. }
    \label{fig:sim lh}
\end{figure*}

The numerical simulation of the SQUID $\Delta$-network coupler is shown in FIG.~\ref{fig:sim lh}(a), (b) and (c). We assume the two side SQUIDs are not biased so $\Phi_{e1}=\Phi_{e2}=0$. The middle flux $\Phi_c$ is swept. We use the full overlap method discussed in Sec.~\ref{sec:overlap method} to extract J and use exact numerical diagonalization to calculate $ZZ$: $ZZ=E_{11}-E_{10}-E_{01}+E_{00}$. For simplicity, we only project the system onto the computational subspace.

We find that by tuning the flux inside the coupling SQUID ($\varphi_c$),  $J_{00}$ can be tuned to zero. This indicates full localization between the two qubits. The residual $ZZ$ interaction can be suppressed below 50 kHz. To find the best trade-off point between $J$ and $ZZ$, we minimize the cost function $|J|+|ZZ|$ to find the optimal off-point. As shown in FIG.~\ref{fig:sim lh}, the optimal flux point is around $\varphi_c=0.2$, giving $J=-18.1 $ kHz and $ZZ=-3.7$ kHz.

The second coupler design we propose is the tunable inductance $\Delta$-network shown in FIG.~\ref{fig:circuit designs l and r}(b). In experiments, tunable linear inductors can be implemented as arrays of SQUIDs. Here we omit the implementation details and model them as purely tunable linear inductors. Similarly to the SQUID $\Delta$-network case, the full circuit Hamiltonian reads
\begin{equation}\label{eq:tunable L H}
\begin{split}
        H=&4\boldsymbol{n}^T\mathbf{E}_C\boldsymbol{n}-E_{Ja}\cos(\varphi_a)-E_{Jb}\cos(\varphi_b)\\
        &+\frac{E_{L1}}{2}\varphi_1^2+\frac{E_{L2}}{2}\varphi_2^2+\frac{E_{L12}}{2}(\varphi_1+\varphi_2)^2   
\end{split}
\end{equation}
where $E_{Lk}=\frac{1}{L_k}\left(\frac{\Phi_0}{2\pi}\right)^2$. Since we assume the inductors to be tunable, this coupler design also supports two tunable coupler modes via $E_{L1}$ and $E_{L2}$. The mode-mode coupling is tunable via $E_{L12}$.
The fact that the Hamiltonian in Eq.~(\ref{eq:tunable L H}) shares the exact same capacitance matrix and $\mathbf{E}_C$ matrix as the SQUID $\Delta$-network in  Eq.~(\ref{eq:3SQUID H}) is because the two coupler designs have the same capacitance network and topology. The only difference is that the SQUIDs are replaced by linear inductors. We use the parameters in TABLE.~\ref{tab:3t} in numerical simulations. The numerical results are shown in FIG.~\ref{fig:sim lh}(d), (e) and (f). We find that the tunable inductance design in general performs better than the SQUID design in terms of turning off interactions. Analogous to the SQUID design, the system decouples near the coupler degeneracy point, at which the coupler eigenmodes separate into symmetric and antisymmetric modes. As shown in the plots, the optimal $J$ and $ZZ$ are found to be -5.0 kHz and -0.7 kHz. We attribute this performance improvement to the linearity of the coupler. This observation also highlights a general design principle: linear circuit elements help suppress nonlinearity-induced errors, such as residual $ZZ$ interactions in the system.

\begin{table}[h]
    \centering
    \begin{tabular}{|c|c|}
        \hline
        Parameter & Value \\
        \hline
        $E_{Ja}$ & 14 GHz \\
        \hline
        $E_{Jb}$ & 18 GHz \\
        \hline
        $L_1$ & 30 nH \\
        \hline
        $L_2$ & 25 nH \\
        \hline
        $C_a$ & 70 fF \\
        \hline
        $C_b$ & 70 fF \\
        \hline
        $C_1$ & 70 fF \\
        \hline
        $C_2$ & 80 fF \\
        \hline
        $C_{ab}$ & 0.05 fF \\
        \hline
        $C_{12}$ & 10 fF \\
        \hline
        $C_{a1}$ & 10 fF \\
        \hline
        $C_{b2}$ & 10 fF \\
        \hline
        $C_{gi}$ & 50 fF \\
        \hline
    \end{tabular}
    \caption{Parameters used in the numerical simulation of the tunable inductance $\Delta$-network coupler.}
    \label{tab:3t}
\end{table}

\section{Selective Control of Effective Coupling in Excitation Manifolds and two-qubit gates}\label{sec: selective J}

A key advantage of two-mode couplers is their ability to generate distinct effective $J$ couplings across different excitation subspaces. Of particular relevance is the ability to engineer configurations where the coupling in the one-excitation manifold is strong (e.g., for gate operations), while the coupling in the two-excitation manifold is suppressed, or vice versa. Such selective control enables high-fidelity gates by minimizing leakage into unwanted states. In this section, we show analytical derivations and numerical simulations to demonstrate the selective control of couplings.

\subsection{Effective couplings}

While a full analytical treatment is hindered by transmon nonlinearity, insight can be obtained using a perturbative Schrieffer–Wolff (SW) analysis, which we then validate numerically.
\begin{figure*}[t]
\centering
\includegraphics[width=0.245\textwidth]{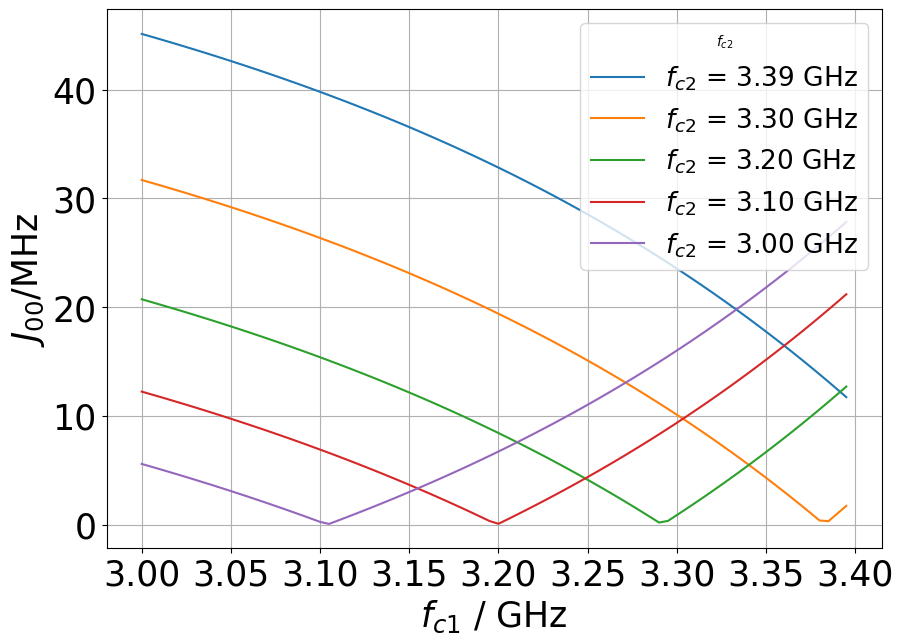}\put(-90,97){\textbf {(a)}}
\includegraphics[width=0.245\textwidth]{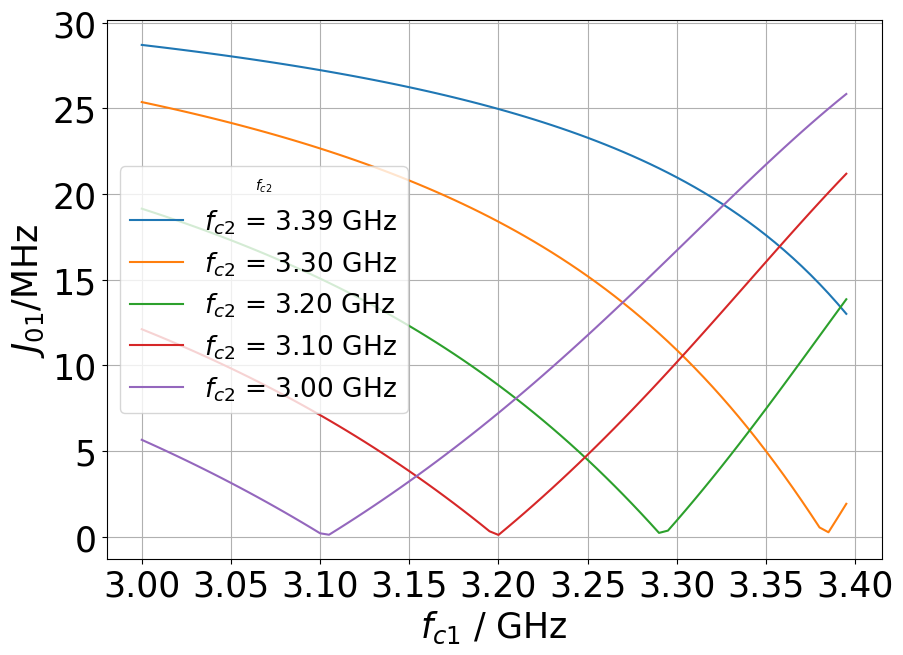}\put(-90,97){\textbf {(b)}}
\includegraphics[width=0.245\textwidth]{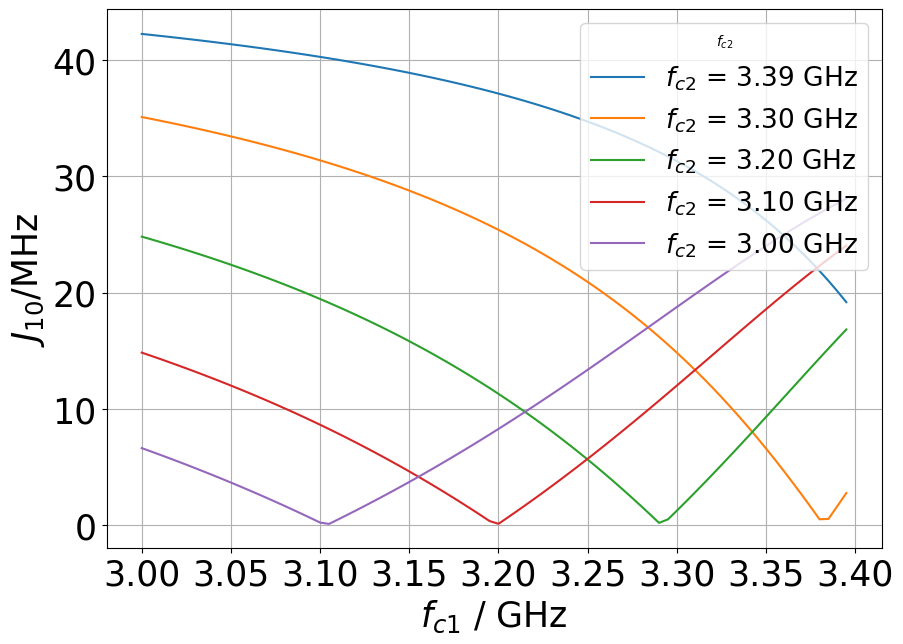}\put(-90,97){\textbf {(c)}}
\includegraphics[width=0.245\textwidth]{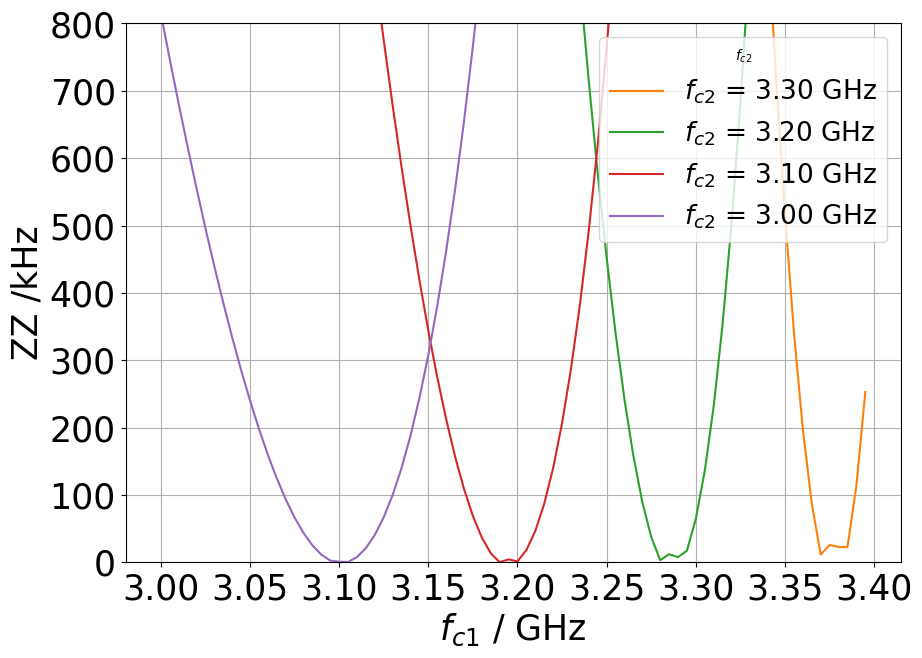}\put(-90,97){\textbf {(d)}}
\caption{\textbf{$J$ couplings and $ZZ$ interaction in the on-mode.} The two transmons are tuned on resonance. In the simulation, we vary frequencies of modes 1 and 2. (a) (b) (c) $J_{00}, J_{01}$ and $J_{10}$. The three couplings in general show similar trends but vary in magnitude. (d) $ZZ$ interaction. $ZZ$ can be suppressed below 100 kHz. At some points, $ZZ < 10$ kHz. }
\label{fig:Js on}
\end{figure*}

\begin{figure}[t]
     \centering
     \includegraphics[width=0.48\textwidth]{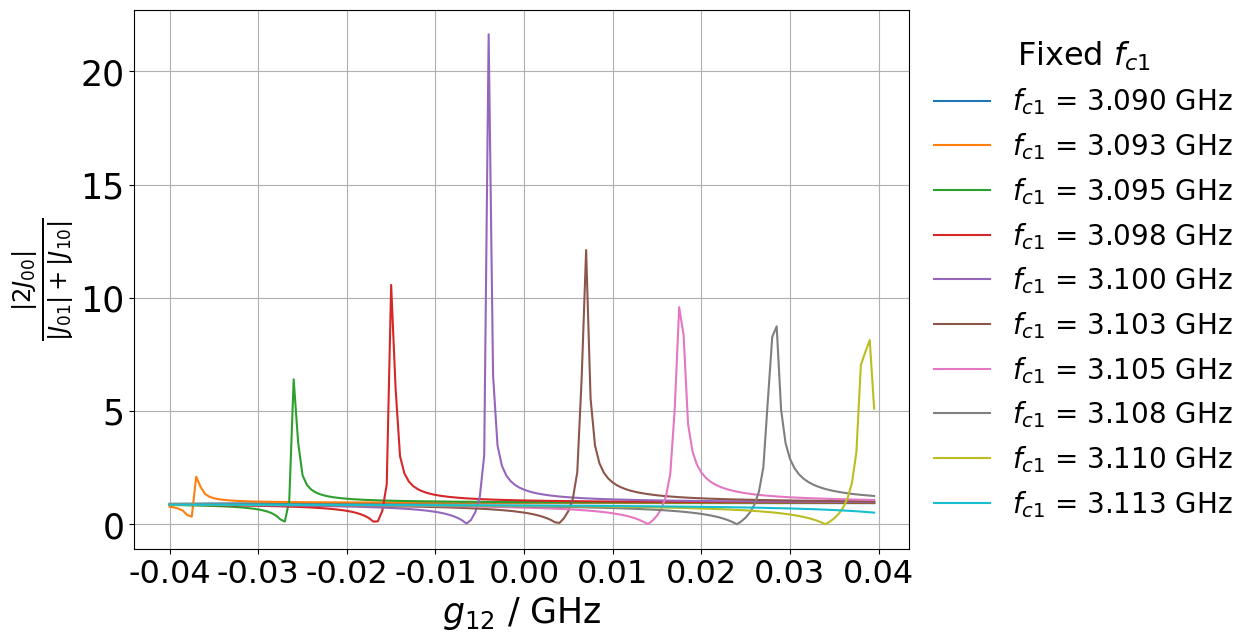}
        \caption{\textbf{The ratio of couplings in the one and two excitation manifolds.} The frequency of the second coupler mode is fixed at 3.0 GHz. The coupling ratio is around 1 in most areas, showing nonlinearity within a narrow window. The spikes and dips are around the zeros of $J_{00}, J_{01}$ and $J_{10}$. In extreme cases, the ratio can exceed 20.}
        \label{fig:ratio}
\end{figure}
Assuming the coupler modes are already diagonalized and denoted by frequencies $\Tilde{\omega}_{1,2}$, the effective exchange coupling between transmons via mode $k$ ($k = 1,2$) in a given excitation subspace is approximately:
\begin{equation}\label{eq:J SW}
\begin{split}
    J_{mn} = g_{ab}-&\frac{\Tilde{g}_{a1} \Tilde{g}_{b1}}{2} \Bigg( \frac{1}{\Tilde{\omega}_1 - \omega_a - m\delta_a} + \frac{1}{\Tilde{\omega}_1 + \omega_a + m\delta_a}\\
    &+\frac{1}{\Tilde{\omega}_1 - \omega_b - n\delta_b}+\frac{1}{\Tilde{\omega}_1 + \omega_b + n\delta_b} \Bigg) \\
    &- \frac{\Tilde{g}_{a2} \Tilde{g}_{b2}}{2} \Bigg( \frac{1}{\Tilde{\omega}_2 - \omega_a - m\delta_a}+\frac{1}{\Tilde{\omega}_2 + \omega_a + m\delta_a} \\
    &+ \frac{1}{\Tilde{\omega}_2 - \omega_b - n\delta_b}+\frac{1}{\Tilde{\omega}_2 + \omega_b + n\delta_b} \Bigg),
\end{split}
\end{equation}
where $J_{mn}$ denotes the effective coupling between states $\ket{m+1, 0, 0, n}$ and $\ket{m, 0, 0, n+1}$.  Specifically, we will focus on couplings in one-excitation and two-excitation manifolds: $J_{00}, J_{01}$ and $J_{10}$, 
which yield:
\begin{widetext}
\begin{align}\label{eq:3 J SW}
\begin{split}
    J_{00} &= \frac{\Tilde{g}_{a1} \Tilde{g}_{b1}}{2} \left( \frac{1}{\Tilde{\omega}_1 - \omega_a}+\frac{1}{\Tilde{\omega}_1 + \omega_a} + \frac{1}{\Tilde{\omega}_1 - \omega_b} +\frac{1}{\Tilde{\omega}_1 + \omega_b} \right)
           + \frac{\Tilde{g}_{a2} \Tilde{g}_{b2}}{2} \left( \frac{1}{\Tilde{\omega}_2 - \omega_a}+\frac{1}{\Tilde{\omega}_2 + \omega_a} + \frac{1}{\Tilde{\omega}_2 - \omega_b}+\frac{1}{\Tilde{\omega}_2 + \omega_b} \right), \\
    J_{01} &= \frac{\Tilde{g}_{a1} \Tilde{g}_{b1}}{2} \left( \frac{1}{\Tilde{\omega}_1 - \omega_a} +\frac{1}{\Tilde{\omega}_1 + \omega_a} + \frac{1}{\Tilde{\omega}_1 - \omega_b - \delta_b} +\frac{1}{\Tilde{\omega}_1 + \omega_b + \delta_b} \right)
           \\
           &+ \frac{\Tilde{g}_{a2} \Tilde{g}_{b2}}{2} \left( \frac{1}{\Tilde{\omega}_2 - \omega_a}+\frac{1}{\Tilde{\omega}_2 + \omega_a} + \frac{1}{\Tilde{\omega}_2 - \omega_b - \delta_b}+\frac{1}{\Tilde{\omega}_2 + \omega_b + \delta_b} \right), \\
    J_{10} &= \frac{\Tilde{g}_{a1} \Tilde{g}_{b1}}{2} \left( \frac{1}{\Tilde{\omega}_1 - \omega_a - \delta_a}+\frac{1}{\Tilde{\omega}_1 + \omega_a + \delta_a} + \frac{1}{\Tilde{\omega}_1 - \omega_b}+\frac{1}{\Tilde{\omega}_1 + \omega_b} \right)\\
           &+ \frac{\Tilde{g}_{a2} \Tilde{g}_{b2}}{2} \left( \frac{1}{\Tilde{\omega}_2 - \omega_a - \delta_a}+\frac{1}{\Tilde{\omega}_2 + \omega_a + \delta_a} + \frac{1}{\Tilde{\omega}_2 - \omega_b}+\frac{1}{\Tilde{\omega}_2 + \omega_b} \right).
\end{split}
\end{align}
\end{widetext}

When the detunings between the coupler modes and transmons are comparable to the transmon anharmonicities, the nonlinear dependence of $J_{01}$ and $J_{10}$ becomes significant. 

To enable destructive interference in higher excitation sectors, the coupler modes are designed to possess opposite parities.  This ensures that the product $\Tilde{g}_{a1} \Tilde{g}_{b1}$ has the opposite sign to $\Tilde{g}_{a2} \Tilde{g}_{b2}$, as previously noted in~\cite{Mundada_2019,Moskalenko_2021}. 
We also require that the coupler modes stay on the same side of the transitions. In our case, we choose: $\tilde{\omega}_{1,2} < \omega_{0\leftrightarrow 1}^{(a,b)} $ and  $\tilde{\omega}_{1,2} > \omega_{1\leftrightarrow 2}^{(a,b)}$ or $\tilde{\omega}_{1,2} < \omega_{1\leftrightarrow 2}^{(a,b)}$. The terms in $J_{01}$ and $J_{10}$ become sign-opposed due to the anharmonic de-tuning shifts, allowing for near-exact cancellation. This configuration thus enables strong $J_{00}$ coupling for entangling gates while suppressing higher-order transitions that are related to energy loss whose statistics correspond  to entanglement suppression error and entropy flow out of the quantum system \cite{Nazarov-parallel,Ansari2015exact,rapp2025}.

In FIG.~\ref{fig:Js on}, we show the simulation of J's while the two qubits are resonant. We choose the bare qubit frequencies: $\omega_a=3.579$ GHz and $\omega_b=3.6$ GHz. This configuration ensures that the qubits are resonant in the dressed basis. Other parameters are the same as in Table~(\ref{tab:para}). 
Since the qubits are now resoannt. The approximated overlap formulas in Appendix.~\ref{appen:J} are not fully valid anymore. We therefore use the exact overlap method as presented in Sec.~\ref{sec:overlap method} to evaluate the coupling in the one-excitation subspace $J_{00}$. To be more precise, we first project the full diagonalization transformation $U$ onto the computational subspace to get the reduced matrix $U_r$ and then perform Singular Value Decomposition (SVD) of $U_r$ to calculate the approximated unitary transformation and effective coupling. The couplings in the two-excitation subspace are evaluated using the approximated formulas in Appendix.~\ref{appen:J} as the states $|1001\rangle$ are not resonant with $|2000\rangle$ or $|0002\rangle$.

All three $J$ couplings and $ZZ$ interaction can reach zero. At first glance, it might feel surprising that even though the two qubits are tuned to resonance, the interactions, especially $J_{00}$ can still be tuned to zero. In fact, this shows the two-mode coupler's ability to fully localize and decouple the qubits, regardless of their frequency proximity. The J couplings are tunable from a few tens of MHz to zero, providing the high on/off ratio essential for high-fidelity operations. Notably, the $ZZ$ interaction can be easily suppressed below 10 kHz in our setup, underscoring the multi-mode coupler’s potential for superior crosstalk mitigation and versatile gate operations. One of our motivations for introducing multi-mode couplers is to achieve nonlinear control of interactions in the one-excitation subspace and the two-excitation subspaces. To characterize this nonlinearity, we plot $\frac{2|J_{00}|}{|J_{01}|+|J_{10}|}$ in FIG. \ref{fig:ratio}. For a linear system, this ratio remains approximately 1. In our simulation, the ratio stays around 1 for most of the parameter space but diverges sharply at specific coupler frequencies. The nonlinear Behavior arises because the three couplings $J_{00},J_{10}$ and $J_{01}$ reach zero at slightly different points. For instance, when $J_{10}$ and $J_{01}$ are nulled while $J_{00}$ remains finite, the ratio peaks.  While various parameters influence these nonlinear dynamics, we show an example of tuning the coupler frequency and the coupler mode-mode coupling, keeping other hardware parameters constant. A comprehensive study on the independent control and optimization of these couplings is required to fully exploit this degree of freedom, which remains beyond the scope of this work.

\subsection{$i$SWAP and CPHASE Gate dynamics}

\begin{figure*}[ht!]
\centering
\includegraphics[width=0.3\textwidth]{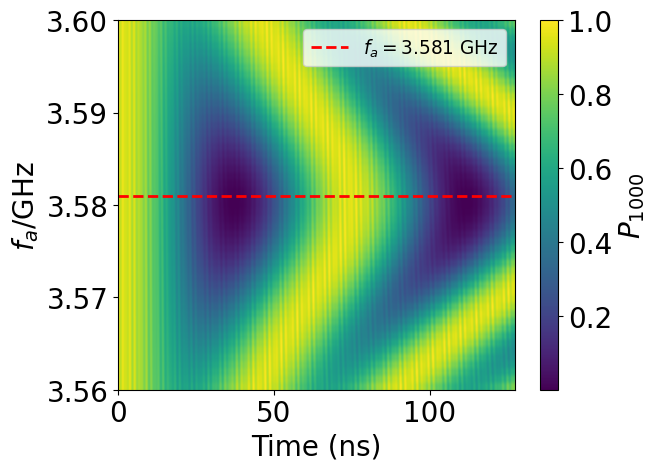}\put(-130,110){\textbf {(a)}} 
\includegraphics[width=0.3\textwidth]{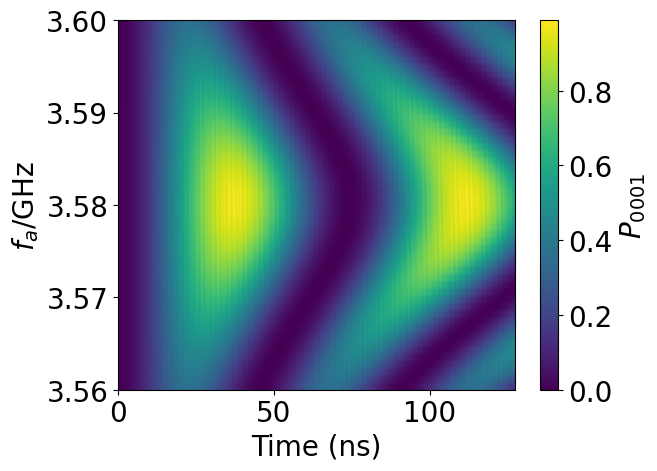}\put(-130,110){\textbf {(b)}}
\includegraphics[width=0.3\textwidth]{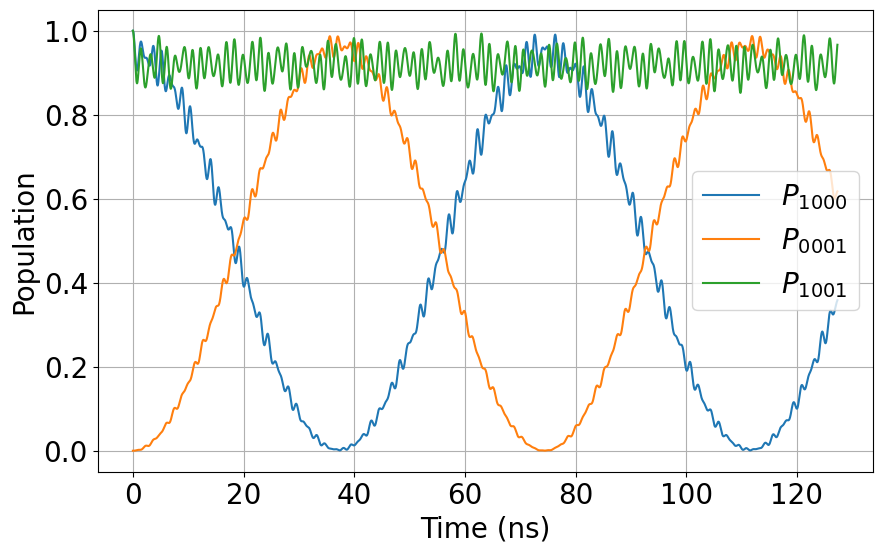}\put(-130,110){\textbf {(c)}}\\
\includegraphics[width=0.3\textwidth]{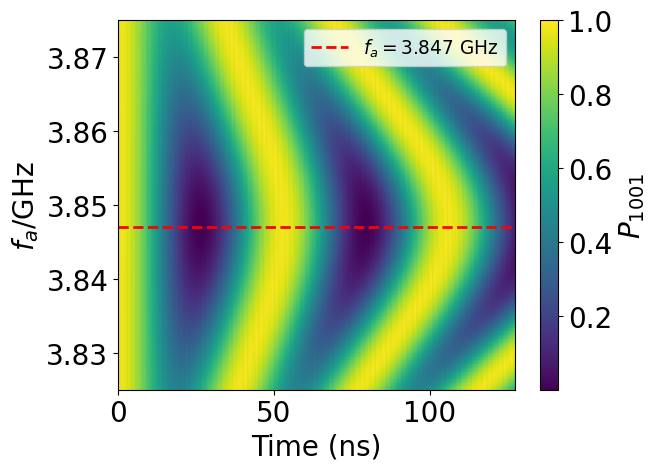}\put(-130,110){\textbf {(d)}} 
\includegraphics[width=0.3\textwidth]{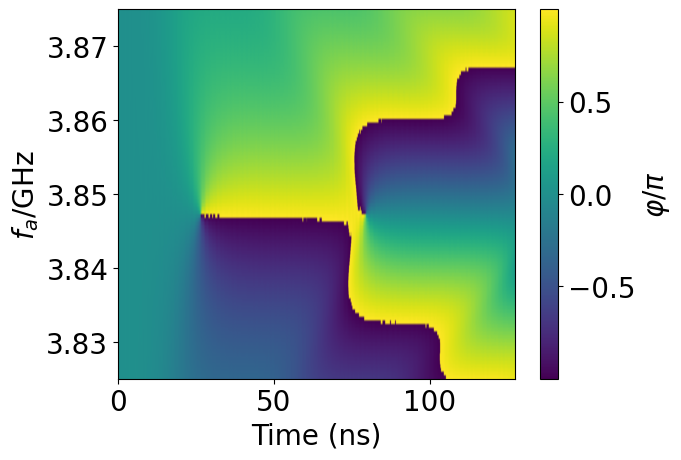}\put(-130,110){\textbf {(e)}}
\includegraphics[width=0.3\textwidth]{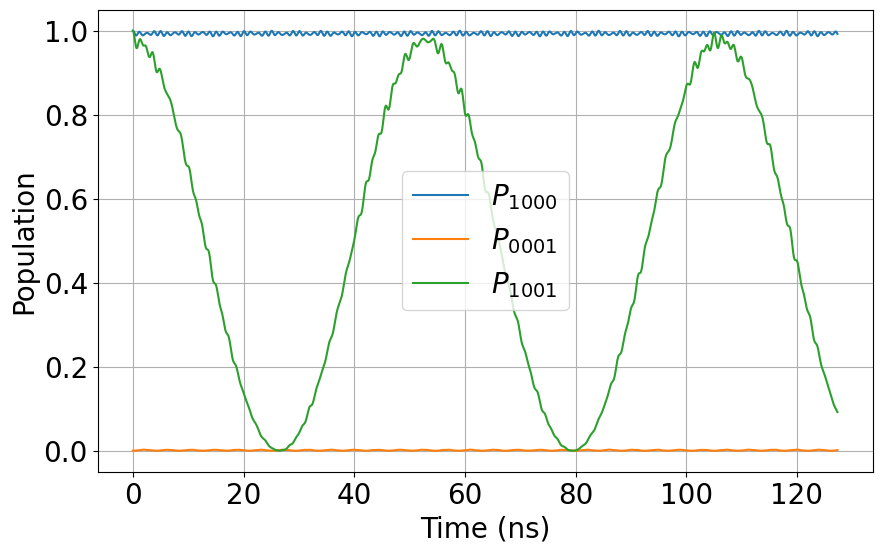}\put(-130,110){\textbf {(f)}}
\caption{\textbf{Real-time simulation of the $i$SWAP gate and the CPHASE gate.} In the simulation we choose the off-point is chosen to be $f_a=4.2 $ GHz,$ f_b= 3.6$ GHz,$ f_1= 3.298$ GHz, and $f_2=3.206$ GHz. We choose the coupler frequencies to be $f_1= 3.2$ GHz and $f_2= 3.0$ GHz to turn the interaction on. The $i$SWAP gate dynamics is shown in (a),(b) and (c). The initial state is $|1000\rangle$. The frequency of transmon $b$ is fixed: $f_b=3.6$ GHz while the frequency of transmon $a$, $f_a$ is swept.  (a) Population of state $|1000\rangle$. Excitation is swapped from $|1000\rangle$ to $|0001\rangle$. The full iSWAP gate is achieved when the frequency of the frist qubit $f_a=3.581$ GHz and at gate time $t_g \approx $ 30 ns. The optimal $f_a$ is marked by a red dashed line. (b) Population of state $|0001\rangle$. Excitation is swapped from $|1000\rangle$ to $|0001\rangle$. (c)The time evolution of the state populations along the full-SWAP line as marked in (a). The population of the state $|1001\rangle$ with the initial state being $|1001\rangle$ is also included to quantify leakage. While the $|1000\rangle$ to $|0001\rangle$ SWAP can be executed with high fidelity, there is some population fluctuation of $|1001\rangle$ due to leakage to coupler modes. This can be mitigated by using a smoother pulse shape and other optimal control techniques. The CHPASE gate dynamics is shown in (d),(e) and (f). The initial state is $|1001\rangle$. The frequency of transmon $b$ is fixed: $f_b=3.6$ GHz while the frequency of transmon $a$, $f_a$ is swept.(d)Population of state $|1001\rangle$. The excitation is coherently swapped from $|1001\rangle$ to $|2000\rangle$ and then back to $|1001\rangle$. The full SWAP coupler frequency is marked by the red dashed line. (e) Conditional phase on $|1001\rangle$. The phase is limited from $-\pi$ to $\pi$. The linecut at $f_a=3.847$ GHz is when the full $\pi$ is achieved. (f)The time evolution of the state populations along the full-SWAP line as marked in (d). The population of the state $|1001\rangle$ completes a full cycle at $t_g \approx 50$ ns. We also plot the population of $|1000\rangle$ and $|0001\rangle$ with the initial state being $|1000\rangle$ to show that the transitions in the one-excitation subspace are largely suppressed.}
\label{fig:iSWAP}
\end{figure*}

Building on the results above, we can show how to operate the iSWAP gate and the CPHASE gate with our two-mode coupler setup. We perform numerical simulations of the time-evolution of the two-transmon-two-coupler-mode system, see FIG.~\ref{fig:iSWAP}. For the $i$SWAP gate, excitation can be coherently swapped between the two transmons by tuning them into resonance and turning on the coupler. For the CPHASE gate, we tune the computational state $|1001\rangle$ on resonance with the uncomputational state $|2000\rangle$. The population on $|1001\rangle$ is swapped to $|2000\rangle$ and then back to $|1001\rangle$. The state $|1001\rangle$ picks up a conditional phase $\varphi =\pi$ after one cycle of evolution.
In both cases, the gate time is around $30-50$ ns.
We find in the $i$SWAP case, there is population fluctuation of computational state $|1001\rangle$ when the initial state is $|1001\rangle$. While the average leakage is approximately $5\%$, reducing baseline gate fidelity, this can be mitigated by adjusting the coupler parameters to align the leakage oscillations such that the $|1001\rangle$ population is recovered when the full $i$SWAP completes. Furthermore, the use of sophisticated pulse-shaping and optimal control tools offers a clear path toward minimizing these non-computational transitions in practical implementations.\cite{PhysRevLett.103.110501,PRXQuantum.5.030353,gao2025ultrafastsinglequbitgates}. The CPHASE gate, on the other hand, does not induce much unwanted transition in the first excitation subspace when the system is initialized at $|1000\rangle$.

We want to point out that what we present here is not a fully optimized gate. Rather, our objective is to introduce a novel coupler architecture that provides superior localization and more versatile control. Our coupler design has three tunable parameters (two coupler mode frequencies and one mode-mode coupling) in contrast of one tunable parameter (coupler frequency) in the standard tunable coupler design. While this increased flexibility introduces a calibration challenge, it simultaneously opens a broader landscape for advanced control schemes. We encourage the community to explore the rich engineering possibilities and physics behind the multi-mode coupler design presented in this work. For example, apart from using standard pulse-shaping techniques to improve $i$SWAP and CPHASE gate fidelity, one can also combine the two gates. Since there are multiple degrees of freedom in the coupler, it is possible to turn on the two transitions simultaneously. Therefore, the control scheme can be easily extended to implement continuous fSim gates\cite{Foxen_2020,jiang2024concurrentfermionicsimulationgate}.

\section{Coupling multiple qubits with a multi-mode coupler}\label{sec:modular QPU}

\begin{figure}[ht!]
     \centering

     \includegraphics[width=0.48\textwidth]{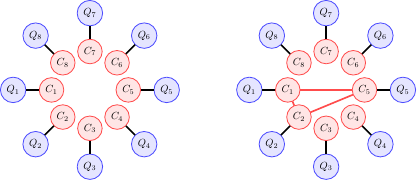}
     \put(-250,100){\textbf {(a)}}
     
     \hfill

         \includegraphics[width=0.48\textwidth]{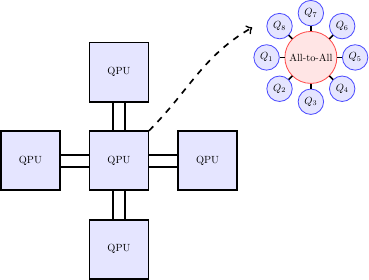}
         \put(-250,120){\textbf {(b)}}
    \caption{\textbf{Coupling multiple qubits with a multi-mode coupler}  (a)All-to-all coupling. An example of turning on interactions among qubits 1,2, and 5. Qubits are initially coupled to one associated coupler mode each. Interactions can be turned on by coupling the corresponding associated coupler modes. (b) A modular QPU design. Qubits are coupled via multi-mode couplers within each module to facilitate all-to-all connectivity.}
        \label{fig:multi q c}

\end{figure}

In this section, we sketch how our multi-mode coupler design can be implemented in experiments to support multi-qubit operations and all-to-all connectivity. Suppose $n$ qubits are all coupled to a common $n$-mode coupler, for instance, a metamaterial ring resonator. Following the general theory in Sec.~\ref{sec:general theo} and the example in Sec.~\ref{sec:full_localization}, we propose the following procedure for tuning the coupler. If we have enough control over each mode, the coupler can be tuned into the off point where each and every qubit is only coupled to its associated coupler mode and there is no mode-mode coupling inside the coupler, see FIG.\ref{fig:multi q c}. At this off point, the whole system is separated into n non-interacting subspaces with each one has one qubit and one coupler mode in it. To selectively turn on qubit-qubit interactions, one can turn on interactions among the associated coupler modes by tuning the coupler. The coupler mode hybridization will mediate qubit interactions corresponding to the associated coupler modes. For experimental realization, we propose two designs, both natural extensions of our SQUID-based and tunable-inductance architectures.
The first, extending the SQUID design, consists of a coupler formed by a loop of SQUIDs with additional tunable SQUIDs inserted in between. The mode frequencies and inter-mode coupling can be controlled via the applied SQUID fluxes.
The second design employs tunable inductances and is analogous to a metamaterial ring resonator\cite{PRXQuantum.5.020325}, but with a tunable inductor in each cell. Here, mismatches in the inductances generate controllable mode-mode coupling. 

This all-to-all coupler is, of course, still localized. Long-distance interaction cannot be achieved in this fashion unless there is a direct link connecting the two parties. We believe that this design is particularly suitable for modular QPU architectures\cite{PhysRevApplied.21.054063,PhysRevX.14.041030,Ihssen_2025}. Within the module, qubits can be connected to a common coupler to enable all-to-all connectivity. Inter-module coupling will be implemented with other long-distance coupling links, as shown in FIG. \ref{fig:multi q c}(b). For example, the inter-module coupling can be implemented by a metamaterial coupler\cite{PRXQuantum.5.020325}.

\section{Summary}

We present a systematic theoretical framework for the design and control of multi-mode couplers in superconducting quantum processors. In contrast to conventional single-mode couplers, our approach enables selective engineering of interactions across different excitation manifolds, thereby offering new routes to suppress leakage and unwanted unitary errors during gate operations. As a concrete demonstration, we analyze in detail a two-transmon system coupled through a two-mode tunable coupler. Numerical simulations show that this architecture can realize $i$SWAP and CPHASE gates  by exploiting independent control of one- and two-excitation subspaces. Furthermore, we provide explicit examples of lumped-element circuit designs that illustrate the practical feasibility of our proposal. Beyond demonstrating a specific gate, our results highlight how multi-mode couplers can combine flexibility, strong tunability, and complete qubit localization at the decoupled point. We believe this work lays the groundwork for the next generation of high-performance couplers in scalable quantum processing units.

\section*{Acknowledgments}
The authors thank David DiVincenzo for fruitful discussions and insightful comments. ZJ and MHA also thank Xuexin Xue for cross-checking the numerical results of the coupler circuits using the software Cirqubit (https://cirqubit.com/).
This work was financed by the Federal Ministry of Research, Technology, and Space (BMFTR) within project QSolid (FKZ: 13N16149 and 13N16151).

\clearpage
\bibliography{sample}

\appendix

\section{Derivation of decoupling with the two-mode coupler using the Lagrangian method }\label{appen:Lag}

Let's assume the system Lagrangian reads:

\begin{equation}\label{eq:setup Lag}
\begin{split}
&L=L_T+L_C+L_{int}\\
&L_C=\frac{1}{2} \dot{\Vec{\phi}}^T_cC\dot{\Vec{\phi}}_c-\frac{1}{2} \Vec{\phi}^T_c\frac{1}{L}\Vec{\phi}_c\\
&L_{int}=\begin{pmatrix}
      \dot{\phi}_a &\dot{\phi}_b
  \end{pmatrix}\begin{pmatrix}
  C_{a1} & C_{a2}  \\ 
 C_{b1} & C_{b2} \\
  \end{pmatrix} \begin{pmatrix}
      \dot{\phi}_1\\
      \dot{\phi}_2
  \end{pmatrix}
\end{split}
\end{equation}

The transmon Lagrangian is not explicitly written out because it does not contain any coupler degree of freedom. The two coupler modes are labeled as $\phi_1$ and $\phi_2$ and the transmon modes are $\phi_a$ and $\phi_b$. We have assumed that the transmons are capacitively coupled to the coupler.

We start by finding the normal modes of the coupler via a transformation $\eta = A \phi_c $. This transformation can be found using different methods. We present one of them. We first introduce an intermediate transformation $x = \sqrt{C} \phi_c $. Under this transformation, the coupler Lagrangian becomes:

\begin{eqnarray}
    L_C=\frac{1}{2} \dot{\Vec{x}}^T\dot{\Vec{x}}-\frac{1}{2} \Vec{x}^TC^{-1/2}\frac{1}{L}C^{-1/2}\Vec{x}
\end{eqnarray}
The normal modes are found by diagonalizing the matrix $C^{-1/2}\frac{1}{L}C^{-1/2}$ with another transformation $\eta=Dx$: $C^{-1/2}\frac{1}{L}C^{-1/2}\eta=\omega^2 \eta$. After the transformations, the coupler Lagrangian contains two uncoupled modes:
\begin{eqnarray}
    L_C=\frac{1}{2} \dot{\Vec{\eta}}^T\dot{\Vec{\eta}}-\frac{1}{2} \Vec{\eta}^T\begin{pmatrix}
  \omega_1^2 & 0  \\ 
 0 & \omega_2^2 \\
  \end{pmatrix} \Vec{\eta}
\end{eqnarray}

The total transformation is:

\begin{equation}
    \eta= A\phi
\end{equation}

where $A=D \sqrt{C}$.

The interaction Lagrangian is also transformed:
\begin{equation}
    L_{int}=\begin{pmatrix}
      \dot{\phi}_a &\dot{\phi}_b
  \end{pmatrix}\begin{pmatrix}
  \Tilde{C}_{a1} & \Tilde{C}_{a2}  \\ 
 \Tilde{C}_{b1} & \Tilde{C}_{b2} \\
  \end{pmatrix} \begin{pmatrix}
      \dot{\eta}_1\\
      \dot{\eta}_2
  \end{pmatrix}
\end{equation}

The new coupling capacitance is related to the old capacitance as:

\begin{equation}
    \begin{pmatrix}
  \Tilde{C}_{a1} & \Tilde{C}_{a2}  \\ 
 \Tilde{C}_{b1} & \Tilde{C}_{b2} \\
  \end{pmatrix} =\begin{pmatrix}
  C_{a1} & C_{a2}  \\ 
 C_{b1} & C_{b2} \\
  \end{pmatrix} A^{-1}
\end{equation}

We have shown that the coupling capacitance seen by the transmons is a function of the coupler circuit parameters ($C$ and $L$). By tuning the coupler parameters, the transformation matrix $A$ is changed. We therefore are effectively tuning the coupling between the transmons and the coupler modes. In reality, it is more natural to assume fixed capacitance and tunable inductance. We thus can control transmon-coupler coupling by tuning the inductance in the coupler. Decoupling can be achieved when:

\begin{equation}
    \Tilde{C}_{a1}=\Tilde{C}_{b2}=0 \text{ or } \Tilde{C}_{a2}=\Tilde{C}_{b1}=0
\end{equation}

In either case, the whole system is split into two subsystems that each contains one transmon and one coupler mode and there is no interaction between the two subsystems.

\section{Applying the overlap J method to the two-mode coupler}\label{appen:J}
We apply our overlap method to calculate the effective coupling $J$ between $|01\rangle$ and $|10\rangle$.
We assume that the full Hamiltonian $H$ can be numerically diagonalized and 
$|01\rangle$ and $|10\rangle$ are well separated from other states so that they can be associated with new eigenstates $|\alpha\rangle$ and $|\beta\rangle$ respectively. The corresponding eigenvalues are $\omega_\alpha$ and $\omega_\beta$. $|\alpha\rangle$ and $|\beta\rangle$ can be numerically found by:

\begin{equation}
\begin{split}
        &|\alpha\rangle=\arg \max|\langle 01 | \psi\rangle|, |\psi\rangle \in \{ \text{eigenvectors of } H \}\\
        &|\beta\rangle=\arg \max|\langle 10 | \psi\rangle|, |\psi\rangle \in \{ \text{eigenvectors of } H \}
\end{split}
\end{equation}

We then project eigenstates $|\alpha\rangle$ and $|\beta\rangle$ onto the subspace spanned by $|a\rangle$ and $|b\rangle$ to get a reduced diagonalization matrix $U_{r}$:

\begin{equation}\label{eq:reduced U}
    U_r=\begin{pmatrix}
  \langle 01 | \alpha\rangle & \langle 01 | \beta\rangle  \\ 
 \langle 10 | \alpha\rangle & \langle 10 | \beta\rangle \\
  \end{pmatrix} 
\end{equation}

The effective coupling strength $J$ is found by requiring it to produce the closest diagonalization matrix to $U_r$. To proceed, we write down the general 2 by 2 unitary transformation $U$:

\begin{equation}
    \begin{split}
        U=&e^{i(r_x \sigma_x +r_y \sigma_y +r_z \sigma_z)}\\
        =&e^{i\theta (n_x \sigma_x +n_y \sigma_y +n_z \sigma_z)}\\
        =&\cos (\theta) + i\sin(\theta) ( n_x \sigma_x +n_y \sigma_y +n_z \sigma_z)\\
        =&\begin{pmatrix}
  \cos (\theta) +i \sin(\theta) n_z & \sin(\theta) (i n_x+ n_y)  \\ 
 \sin(\theta) (i n_x- n_y) & \cos (\theta) -i \sin(\theta) n_z  \\
  \end{pmatrix} 
    \end{split}
\end{equation}

As usual, we have defined: $\theta=\sqrt{r_x^2+r_y^2+r_z^2}, n_x=\frac{r_x}{\theta}, n_y=\frac{r_y}{\theta}$ and $ n_z=\frac{r_z}{\theta}$. $r_x, r_y$ and $r_z$ are determined by minimizing the distance between the reduced diagonalization matrix $U_r$ and $U$:
\begin{equation}\label{eq:rs}
    r_x,r_y,r_z=\arg \min |U_r-U|
\end{equation}

The effective Hamiltonian $H_{eff}$ can be calculated from $U, \omega_\alpha$ and $\omega_\beta$ as:
\begin{widetext}
    \begin{equation}
    \begin{split}
        H_{\mathrm{eff}}=&U (-\frac{\omega_\alpha-\omega_\beta}{2}\sigma_z) U^\dag\\
        =&-\frac{\omega_\alpha-\omega_\beta}{2}U  \sigma_zU^\dag\\
        =&-\frac{\omega_\alpha-\omega_\beta}{2} [ (\cos^2(\theta)+(-n_x^2-n_y^2+n_z^2)\sin^2(\theta))\sigma_z+n_x(\sin(2\theta)+2\sin^2(\theta) n_z)\sigma_x\\
        &+n_y(-\sin(2\theta)+2\sin^2(\theta) n_z)\sigma_y]
    \end{split}
\end{equation}

\end{widetext}

$J^\mathrm{eff}$ between $|01\rangle$ and $|10\rangle$ can now be read from the effective Hamiltonian:

\begin{equation}\label{eq:J eff}
\begin{split}
    J^\mathrm{eff}=&-\frac{\omega_\alpha-\omega_\beta}{2}(n_x(\sin2\theta+2n_z\sin^2\theta )\\
    &\pm i n_y(-\sin2\theta+2n_z\sin^2\theta ))
\end{split}
\end{equation}

Assuming $n_y=n_z=0$ and $\theta\rightarrow 0$, we can get the perturbative expansion:

\begin{eqnarray}\label{eq:J pert}
    J^{\mathrm{eff}}=-(\omega_\alpha-\omega_\beta)\sin(\theta)
\end{eqnarray}

In the perturbative regime, where $|\omega_\alpha-\omega_\beta|\gg 0$, the calculation can be further simplified. In fact, according to the localization picture, we can simply approximate $\sin(\theta)$ by the average overlap:

\begin{equation}
    |\sin(\theta)|\approx \sqrt{\frac{|\langle 01 | \beta\rangle|^2+|\langle 10 | \alpha\rangle|^2}{2}}
\end{equation}
The localization-consistent J becomes
\begin{eqnarray}\label{eq:J pert final}
    |J^{\mathrm{eff}}|\approx|\omega_a-\omega_b|\sqrt{\frac{|\langle 01 | \beta\rangle|^2+|\langle 10 | \alpha\rangle|^2}{2}}
\end{eqnarray}
where $\omega_a$ and $\omega_b$ are bare qubit frequencies.

When calculating the effective couplings between $|11\rangle$ and $|02\rangle$ or $|20\rangle$, we replace Eq.~(\ref{eq:J pert final}) with alternative heuristic formulas. Assuming the two transmons have anharmonicity $\delta_1$ and $\delta_2$ and  the dressed $|11\rangle$ state is identified as $|\gamma\rangle$ according to
\begin{equation}
    |\gamma\rangle=\arg \max|\langle 11 | \psi\rangle|, |\psi\rangle \in \{ \text{eigenvectors of } H \}
\end{equation}

The effective couplings can be derived from wavefunction overlaps as

\begin{equation}\label{eq: effe J pertur 11}
    \begin{split}
        &|J_{10}^{\mathrm{eff}}|\approx|\omega_a-\omega_b-\delta_a||\langle20|\gamma\rangle|\\
        &|J_{01}^{\mathrm{eff}}|\approx|\omega_a-\omega_b+\delta_b||\langle02|\gamma\rangle|
    \end{split}
\end{equation}

Eq.~(\ref{eq:J pert final}) and Eq.~(\ref{eq: effe J pertur 11}) are used in numerical simulation for the off-mode. It is very clear from Eq.~(\ref{eq:J pert final}) and Eq.~(\ref{eq: effe J pertur 11}) that when $J^{\text{eff}}$ is zero, the states are localized.

\section{Search for minimal $J_{00}$}\label{appen:mini J}

\begin{figure}
    \centering
    \includegraphics[width=0.45\textwidth]{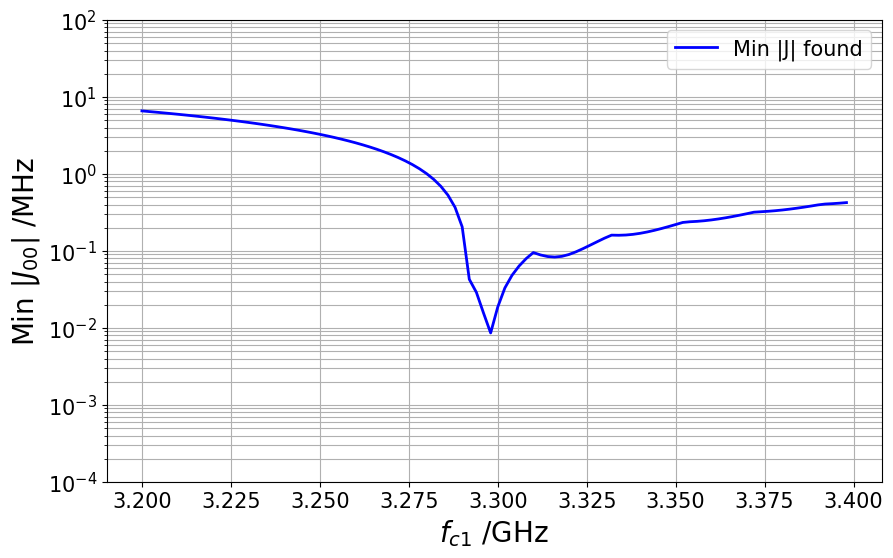}
    \caption{\textbf{Minimal $J_{00}$ found in the off-mode.} Frequencies of the two coupler modes $f_{c1}$ and $f_{c2}$ are swept between $3.2$ GHz and $3.4$ GHz. We numerically search for minimal coupling in the one excitation manifold $J_{00}$ for each $f_{c1}$. The minimal $J_{00}$ can be suppressed below $10$ kHz around $f_{c1}=3.298$ GHz and $f_{c2}=3.206$ GHz.}
    \label{fig:mini J}
\end{figure}

To further show the two-mode coupler's ability to realize localized decoupling, we search for the global minimum of the one-excitation manifold interaction $J_{00}$ by sweeping the coupler mode frequencies $f_{c1}$ and $f_{c2}$. The found minimal $J_{00}$ is shown in FIG.~\ref{fig:mini J}. $J_{00}$ can be suppressed below 100 kHz within a window around $f_{c1}=3.3$ GHz. The global minimal $J_{00}$ is below 10 kHz around $f_{c1}=3.298$ GHz.

\section{Coupler circuit designs}\label{appen:circ}

In this section, we present analytical derivations of the Hamiltonians for the two circuit designs shown in FIG.~\ref{fig:circuit designs l and r}.

\subsection{The SQUID $\Delta$-network design}\label{sec:lh}

\begin{widetext}
\begin{figure*}[ht!]
     \centering
     \includegraphics[width=0.9\textwidth]{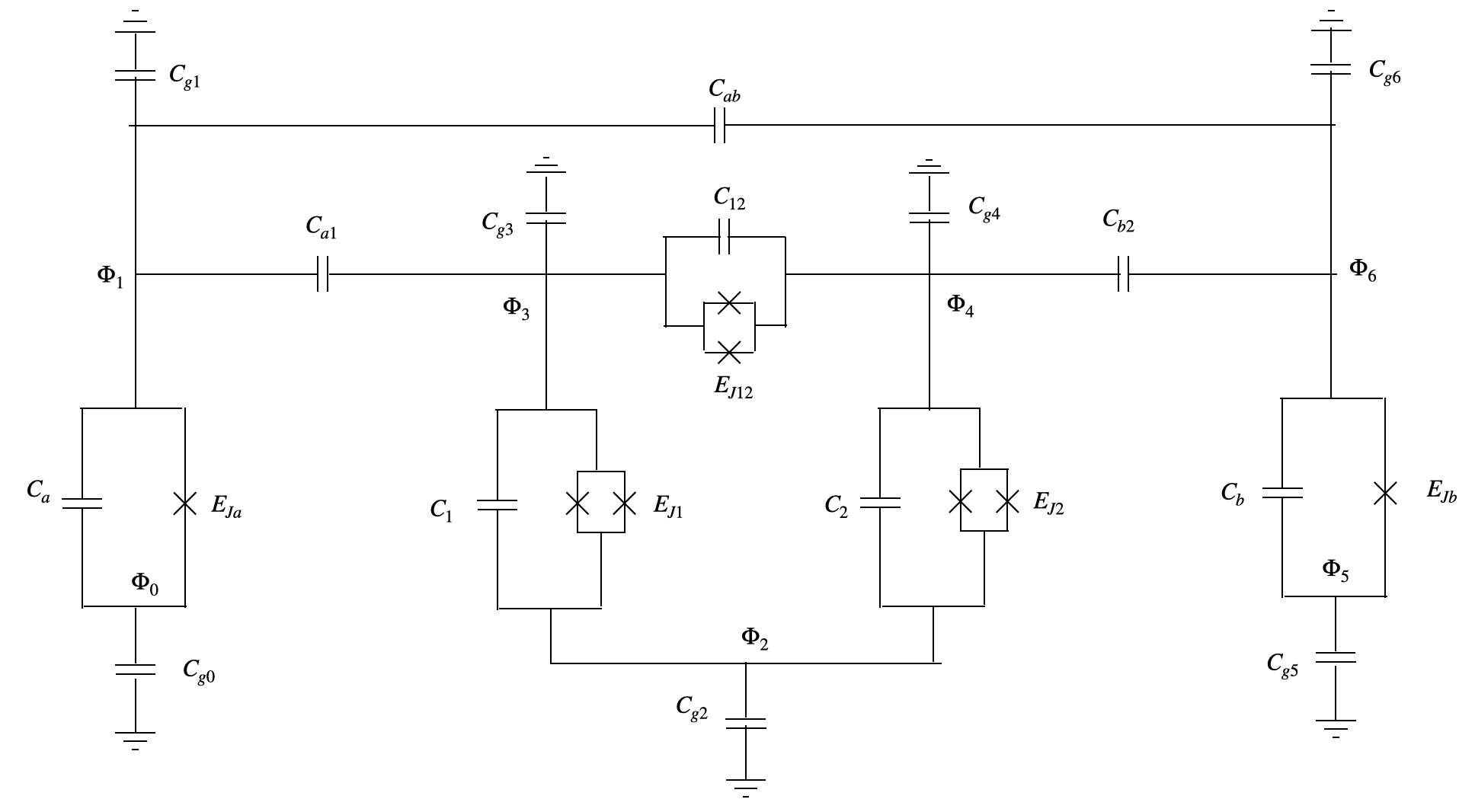}
        \caption{\textbf{The SQUID design.} Node fluxes are labeled as $\Phi_k$. The circuit topology includes a loop encompassing nodes $\Phi_2$, $\Phi_3$, and $\Phi_4$. Although the threading external flux of the $\Phi_2-\Phi_3-\Phi_4$ loop can be distributed across the three SQUIDs, we set the external flux to zero here to simplify the model.}
        \label{fig:squid deri}
\end{figure*}

Following the label in FIG.~\ref{fig:squid deri}, we can write down the capacitance matrix:

\begin{equation}
\mathbf{C} = 
\begin{pmatrix}
C_{g0}+C_a & -C_{a} & 0 & 0 & 0 & 0 & 0 \\
-C_{a} & C_{a}+C_{g1}+C_{a1} & 0 & -C_{a1} & 0 & 0 & 0 \\
0 & 0 & C_{g2}+C_1+C_2 & -C_1 & -C_2 & 0 & 0 \\
0 & -C_{a1} & -C_1 & C_{g3}+C_{a1}+C_1+C_{12} & -C_{12} & 0 & 0 \\
0 & 0 & -C_2 & -C_{12} & C_{g4}+C_{12}+C_{2}+C_{b2} & 0 & -C_{b2} \\
0 & 0 & 0 & 0 & 0 & C_{g5}+C_b & -C_b \\
0 & 0 & 0 & 0 & -C_{b2} & -C_b & C_{g6}+C_{b2}+C_b
\end{pmatrix}
\end{equation}
\end{widetext}

The Lagrangian reads

\begin{equation}
\begin{split}
        \mathcal{L}=&\frac{1}{2}\dot{\boldsymbol{\Phi}}^T\mathbf{C}\dot{\boldsymbol{\Phi}}+E_{Ja}\cos(\frac{2\pi}{\Phi_0}(\Phi_1-\Phi_0))\\
        &+E_{Jb}\cos(\frac{2\pi}{\Phi_0}(\Phi_6-\Phi_5))\\
        &+2E_{J1}\cos(\frac{\pi}{\Phi_0}\Phi_{e1})\cos(\frac{2\pi}{\Phi_0}(\Phi_3-\Phi_2))\\
        &+2E_{J2}\cos(\frac{\pi}{\Phi_0}\Phi_{e2})\cos(\frac{2\pi}{\Phi_0}(\Phi_4-\Phi_2)) \\
        & +2E_{J12}\cos(\frac{\pi}{\Phi_0}\Phi_c)\cos(\frac{2\pi}{\Phi_0}(\Phi_4-\Phi_3))
\end{split}
\end{equation}
where $\Phi_{e1},\Phi_{e2},\Phi_{ec}$ are applied through the SQUID loops with junction energy $E_{J1},E_{J2},E_{J12}$ respectively.

Introducing common modes and differential modes, we apply the transformation $S$:
\begin{equation}
    S=\begin{pmatrix}
-1/2 & 0 & 0 & 0 & 1 & 0 & 0 \\
1/2 & 0 & 0 & 0 & 1 & 0 & 0 \\
0 & 0 & 1/3 & -1/3 & 0 & 0 & 1 \\
0 & 0 & -2/3 & -1/3 & 0 & 0 & 1 \\
0 & 0 & 1/3 & 2/3 & 0 & 0 & 1 \\
0 & 0 & 0 & 0 & -1/2 & 1& 0 \\
0 & 0 & 0 & 0 & 1/2 & 1 & 0
\end{pmatrix}
\end{equation}
The modes are related by
\begin{equation}
\mathbf{\Phi}=S\tilde{\mathbf{\Phi}}
\end{equation}
with $\mathbf{\Phi}^T=(\Phi_0,
\Phi_1,\Phi_2,\Phi_3,\Phi_4,\Phi_5,\Phi_6)^T$ and $\tilde{\mathbf{\Phi}}^T=(\Phi_a,
\Phi_b,\Phi_{c1},\Phi_{c2},\Phi_{sa},\Phi_{sb},\Phi_{sc})^T$. $\Phi_{sa},\Phi_{sb},\Phi_{sc}$ are the common modes that will be removed. To proceed, we first write down the Lagrangian in the new basis:
\begin{equation}
\begin{split}
        \mathcal{L}=&\frac{1}{2}\dot{\tilde{\boldsymbol{\Phi}}}^T S^T\mathbf{C}S\dot{\tilde{\boldsymbol{\Phi}}}+E_{Ja}\cos(\frac{2\pi}{\Phi_0}\Phi_a)\\
        &+E_{Jb}\cos(\frac{2\pi}{\Phi_0}\Phi_b)\\
        &+2E_{J1}\cos(\frac{\pi}{\Phi_0}\Phi_{e1})\cos(\frac{2\pi}{\Phi_0}\Phi_{c1})\\
        &+2E_{J2}\cos(\frac{\pi}{\Phi_0}\Phi_{e2})\cos(\frac{2\pi}{\Phi_0}\Phi_{c2}) \\
        & +2E_{J12}\cos(\frac{\pi}{\Phi_0}\Phi_c)\cos(\frac{2\pi}{\Phi_0}(\Phi_{c1}+\Phi_{c2}))
\end{split}
\end{equation}
Note that the three common modes $\Phi_{sa},\Phi_{sb},\Phi_{sc}$ do not show up in the potential. We can eliminate them by using the Euler-Lagrange equation:

\begin{equation}
    \frac{d}{dt}\frac{\partial \mathcal{L}}{\partial \dot{\Phi}_k}-\frac{\partial \mathcal{L}}{\partial \Phi_k}=0
\end{equation}
For the common modes $\Phi_{sa},\Phi_{sb},\Phi_{sc}$, we have $\frac{\partial \mathcal{L}}{\partial \Phi_k}=0$. Therefore $\frac{d}{dt}\frac{\partial \mathcal{L}}{\partial \dot{\Phi}_k}=0$. Assuming the integration constants can be ignored, we arrive at
\begin{equation}
    \frac{\partial \mathcal{L}}{\partial \dot{\Phi}_k}=0 , \quad k=sa,sb,sc
\end{equation}
Let us rewrite the capacitance matrix in the new basis  $S^T\mathbf{C}S$:
\begin{equation}
    S^T\mathbf{C}S=\begin{pmatrix}
        \mathbf{C}_d & \mathbf{C}_{cd}^T\\
        \mathbf{C}_{cd} & \mathbf{C}_{c}
    \end{pmatrix}
\end{equation}
where the capacitance matrix in split into the differential mode part $\mathbf{C}_{c}$, the common mode part $\mathbf{C}_{d}$ and the coupling between them $\mathbf{C}_{cd}$. The Euler-Lagrange equation with respect to the common modes then implies

\begin{equation}
    \mathbf{C}_{cd}\begin{pmatrix}
        \Phi_a\\
        \Phi_b\\
        \Phi_{c1}\\
        \Phi_{c2}
    \end{pmatrix}+ \mathbf{C}_{c}\begin{pmatrix}
        \Phi_{sa}\\
        \Phi_{sb}\\
        \Phi_{sc}
    \end{pmatrix}=0
\end{equation}
We can then express the common modes using differential modes:
\begin{equation}
     \begin{pmatrix}
        \Phi_{sa}\\
        \Phi_{sb}\\
        \Phi_{sc}
    \end{pmatrix}=-\mathbf{C}_{c}^{-1}\mathbf{C}_{cd}\begin{pmatrix}
        \Phi_a\\
        \Phi_b\\
        \Phi_{c1}\\
        \Phi_{c2}
    \end{pmatrix}
\end{equation}
Thus, the kinetic energy term can be transformed as

\begin{equation}
    \frac{1}{2}\dot{\tilde{\boldsymbol{\Phi}}}^T S^T\mathbf{C}S\dot{\tilde{\boldsymbol{\Phi}}}=\frac{1}{2}(\Phi_a, \Phi_b, \Phi_{c1}, \Phi_{c2})\mathbf{C}_{\text{eff}}\begin{pmatrix}
        \Phi_a\\
        \Phi_b\\
        \Phi_{c1}\\
        \Phi_{c2}
    \end{pmatrix}
\end{equation}
with 
\begin{equation}
    \mathbf{C}_{\text{eff}}=\mathbf{C}_d-\mathbf{C}_{cd}^T
        \mathbf{C}_{c}^{-1}\mathbf{C}_{cd}
\end{equation}
We thus have the Lagrangian with the four independent variables:
\begin{equation}
    \begin{split}
        \mathcal{L}=&\frac{1}{2}(\Phi_a, \Phi_b, \Phi_{c1}, \Phi_{c2})\mathbf{C}_{\text{eff}}\begin{pmatrix}
        \Phi_a\\
        \Phi_b\\
        \Phi_{c1}\\
        \Phi_{c2}
    \end{pmatrix}+E_{Ja}\cos(\frac{2\pi}{\Phi_0}\Phi_a)\\
        &+E_{Jb}\cos(\frac{2\pi}{\Phi_0}\Phi_b)\\
        &+2E_{J1}\cos(\frac{\pi}{\Phi_0}\Phi_{e1})\cos(\frac{2\pi}{\Phi_0}\Phi_{c1})\\
        &+2E_{J2}\cos(\frac{\pi}{\Phi_0}\Phi_{e2})\cos(\frac{2\pi}{\Phi_0}\Phi_{c2}) \\
        & +2E_{J12}\cos(\frac{\pi}{\Phi_0}\Phi_c)\cos(\frac{2\pi}{\Phi_0}(\Phi_{c1}+\Phi_{c2}))
\end{split}
\end{equation}
Replacing the indices $c_1, c_2$ by $1,2$ and $\mathbf{C}_{\text{eff}}$ by$\mathbf{C}$, we recover the Lagrangian in Eq.~(\ref{eq:3squid L}).
The Hamiltonian can be calculated as usual:
\begin{equation}
\begin{split}
     H=&\frac{1}{2}\mathbf{Q}^T \mathbf{C}^{-1}\mathbf{Q}+E_{Ja}\cos(\frac{2\pi}{\Phi_0}\Phi_a)\\
        &-E_{Jb}\cos(\frac{2\pi}{\Phi_0}\Phi_b)\\
        &+2E_{J1}\cos(\frac{\pi}{\Phi_0}\Phi_{e1})\cos(\frac{2\pi}{\Phi_0}\Phi_{c1})\\
        &-2E_{J2}\cos(\frac{\pi}{\Phi_0}\Phi_{e2})\cos(\frac{2\pi}{\Phi_0}\Phi_{c2}) \\
        & -2E_{J12}\cos(\frac{\pi}{\Phi_0}\Phi_c)\cos(\frac{2\pi}{\Phi_0}(\Phi_{c1}+\Phi_{c2}))
\end{split}
\end{equation}

\subsection{The tunable inductance design}\label{sec:rh}
Since the tunable inductance design shares the exact same capacitance network and the topology as the SQUID design, the derivation of Lagrangian and Hamiltonian is very similar. Here we directly give the results. The node labels are shown as in FIG.~\ref{fig:tl deri}.
\begin{widetext}
    \begin{figure*}[ht!]
     \centering
     \includegraphics[width=0.9\textwidth]{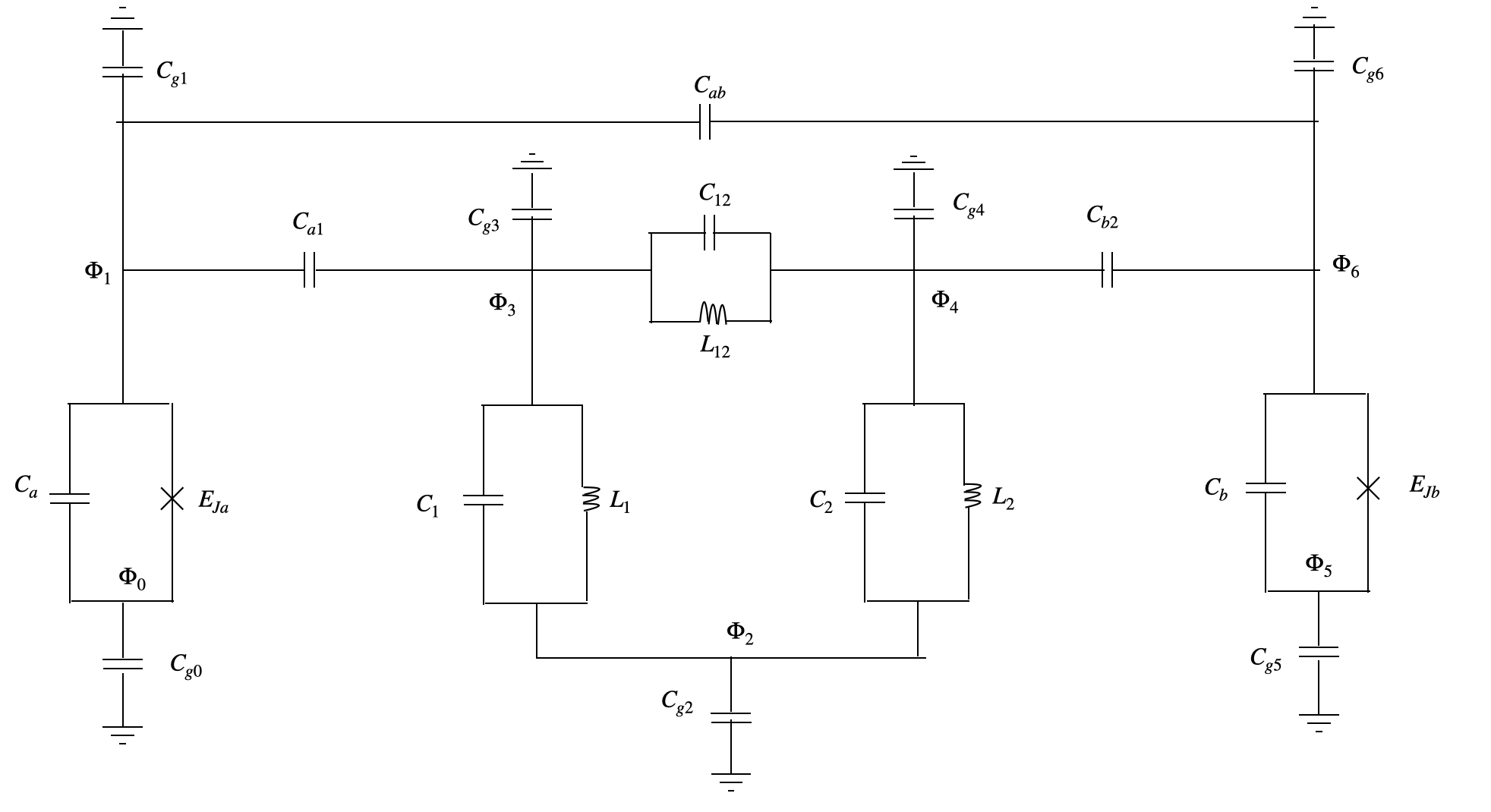}
        \caption{\textbf{The tunable inductance design.} Node fluxes are labeled as $\Phi_k$. As described in the SQUID case, the external flux threading the $\Phi_2-\Phi_3-\Phi_4$ loop is set to zero. }
        \label{fig:tl deri}
\end{figure*}
The Lagrangian is found as

\begin{equation}
    \begin{split}
        \mathcal{L}=&\frac{1}{2}(\Phi_a, \Phi_b, \Phi_{c1}, \Phi_{c2})\mathbf{C}_{\text{eff}}\begin{pmatrix}
        \Phi_a\\
        \Phi_b\\
        \Phi_{c1}\\
        \Phi_{c2}
    \end{pmatrix}+E_{Ja}\cos(\frac{2\pi}{\Phi_0}\Phi_a)\\
        &+E_{Jb}\cos(\frac{2\pi}{\Phi_0}\Phi_b)-\frac{\Phi_{c1}^2}{2L_1}-\frac{\Phi_{c2}^2}{2L_2} -\frac{(\Phi_{c1}+\Phi_{c2})^2}{2L_12}
\end{split}
\end{equation}
The Hamiltonian is 
\begin{equation}
\begin{split}
     H=&\frac{1}{2}\mathbf{Q}^T \mathbf{C}^{-1}\mathbf{Q}+E_{Ja}\cos(\frac{2\pi}{\Phi_0}\Phi_a)\\
        &-E_{Jb}\cos(\frac{2\pi}{\Phi_0}\Phi_b)+\frac{\Phi_{c1}^2}{2L_1}+\frac{\Phi_{c2}^2}{2L_2}+ \frac{(\Phi_{c1}+\Phi_{c2})^2}{2L_{12}}
\end{split}
\end{equation}
\end{widetext}

\end{document}